\documentclass[a4paper,11pt]{article}
\pdfoutput=1 

\usepackage{jheppub} 

\usepackage[T1]{fontenc} 
\usepackage{units}
\usepackage{soul}

\RequirePackage[dvipsnames]{xcolor}

\newcommand{\pt}{p_\perp}
\newcommand{\kt}{k_\perp}

\title{\boldmath Improved background subtraction and a fresh look at jet sub-structure in JEWEL}

\author[a,b]{José Guilherme Milhano}
\author[c,d]{Korinna Zapp}

\affiliation[a]{LIP, Av. Prof. Gama Pinto, 2, P-1649-003 Lisboa, Portugal}
\affiliation[b]{Departmento de F\'isica, Instituto Superior Técnico (IST), Universidade de Lisboa,  Av. Rovisco Pais 1, P-1049-001 Lisboa, Portugal}
\affiliation[c]{Dept. of Astronomy and Theoretical Physics, Lund University, Sölvegatan 14A, S-223 62 Lund, Sweden}
\affiliation[d]{Faculty of Science and Technology, University of Stavanger, Kristine Bonnevies vei 22,\\ 4021 Stavanger, Norway}

\emailAdd{gmilhano@lip.pt}
\emailAdd{korinna.zapp@thep.lu.se}

\abstract{
Interactions of hard partons in the Quark Gluon Plasma (QGP) created with relativistic heavy ion collisions lead to characteristic modifications of the internal structure of reconstructed jets. A large part of the observed jet sub-structure modifications stem from the QGP's response to energy and momentum deposited by hard partons.
Good control over medium response in theoretical calculations is thus instrumental to a quantitative understanding of medium modified (quenched) jets in heavy ion collisions. We present an improved way of handling the medium response in the jet quenching model \textsc{Jewel} and present results for a variety of jet sub-structure observables. The new recoil handling is more versatile and robust than the old scheme, giving a better control over many observables and, in particular, greatly improves the description of the jet mass.
}

\preprint{LU-TH 22-48, MCNET-22-12}

\begin{document} 
\maketitle
\flushbottom

\section{Introduction}
\label{sec:intro}

Jets, collimated bunches of hadrons that result from the branching of energetic quarks or gluons and the subsequent hadronization of the fragments, are pivotal objects in collider physics. They provide a proxy for the fundamental degrees of freedom of Quantum Chromodynamics (QCD) and their study allows for stringent tests of the perturbative calculability of QCD dynamics. The precision understanding of their rich internal structure, usually referred to as jet sub-structure, is instrumental for tests of the Standard Model and in searches for physics beyond it \cite{Nachman:2022emq,Abdesselam:2010pt,Altheimer:2012mn,Altheimer:2013yza,Adams:2015hiv,Larkoski:2017jix,Marzani:2019hun,Kogler:2021kkw}. 

In a heavy-ion collision the branching of energetic partons occurs in the presence of a hot, dense and coloured Quark Gluon Plasma (QGP). The interaction between partons in the branching sequence and the QGP results in modifications of both the parton branching pattern and of the QGP they propagate through (see \cite{Mehtar-Tani:2013pia,Qin:2015srf,Connors:2017ptx,Apolinario:2022vzg} for reviews).
While all modifications arise from the same underlying dynamics, the exchange of colour and energy-momentum between partons and the QGP, it is helpful to describe qualitatively the classes of modifications as follows.  

A first class relates to the ability of interactions to alter the dynamics of the branching sequence through modification of each branching step \cite{Baier:1994bd,Baier:1996sk,Baier:1996kr,Baier:1998yf,Zakharov:1996fv,Zakharov:1997uu,Wiedemann:2000za,Gyulassy:2000er,Guo:2000nz,Wang:2001ifa,Arnold:2000dr,Arnold:2001ms,Arnold:2002ja,Armesto:2011ht,Apolinario:2014csa,Sievert:2019cwq,Mehtar-Tani:2019ygg,Andres:2020vxs,Barata:2020sav,Andres:2020kfg,Barata:2021wuf,Isaksen:2022pkj} and to disrupt the coherence properties of the branching sequence \cite{Mehtar-Tani:2010ebp,Casalderrey-Solana:2011ule,Mehtar-Tani:2011hma,Mehtar-Tani:2012mfa,Casalderrey-Solana:2015bww,Arnold:2015qya,Arnold:2016kek,Arnold:2016mth,Arnold:2016jnq,Dominguez:2019ges,Arnold:2020uzm,Arnold:2021pin,Barata:2021byj,Arnold:2022epx}, thus altering  the relation between subsequent branchings.  

A second class of modifications arise from transfer (loss) of energy-momentum from partons to QGP which may eventually result in a parton becoming equilibrated with the QGP \cite{Iancu:2015uja}, and with the deviation of partons from their original trajectories away from the direction of the branching sequence \cite{Casalderrey-Solana:2010bet}.

These modifications will result in changes, relative to a case where no QGP is present, of what is reconstructed as a jet. Jet reconstruction involves the specification of a set of rules and parameters, a jet algorithm, dictating which particles are to be combined into a specific jet and how their momenta is to be combined. The anti-$\kt$ algorithm \cite{Cacciari:2008gp}, typically used in hadron colliders, recursively clusters particles according to a distance measure that re-weights the distance in the rapidity-azimuth plane with the inverse of the transverse momentum of the hardest particle in a pair thus privileging the early clustering of pairs involving at least a hard particle. Clustering stops at a specified maximum distance, commonly referred to as the jet radius.

Modifications of jets arise from both the loss of particles from the jet reconstruction catchment, for particles radiated to large angles due to QGP effects or particles deviated from their original trajectory, and from the change of particle distribution within the jet.

The modification of the QGP by the traversing partons alters the hydrodynamical evolution of the QGP. A joint description of parton branching and QGP modifications has been explored in \cite{Chen:2017zte,Tachibana:2017syd}. However, for the evolution of the QGP these effects have been argued to be small \cite{Floerchinger:2014yqa}. Due to momentum conservation, the modifications of the QGP are strongest in the vicinity of hard jets. A far more phenomenologically important consequence of the disturbance of the QGP by the traversing partons is thus that this part of this disturbance will be reconstructed as part of a jet. While these QGP response contributions to jets are generically expected to be small, they have been shown \cite{Neufeld:2011yh,1402.6469,He:2015pra,Tachibana:2015qxa,Wang:2013cia,Cao:2016gvr,Casalderrey-Solana:2016jvj,KunnawalkamElayavalli:2017hxo} to be responsible for strong modifications of jet substructure observables.
As these observables have become central to current studies of QGP with jets \cite{Andrews:2018jcm},  an accurate description of this class of jet modifications is an essential part of the study jets in heavy ion collisions.

In the high occupancy environment of a heavy-ion collision, jet reconstruction is only possible in combination with a background subtraction procedure. Background refers to all in an event, specifically all that overlaps with the reconstruction region of a jet, that is unrelated with the jet. In practice, the background to be subtracted is estimated from events with no jets or from 
regions of the same event where no jets are present. A perfect background subtraction would thus result in the elimination of all components of the event uncorrelated with the jet from the final reconstructed jet. One should, however, note that no background subtraction will, nor should, remove QGP response as that part of the QGP is, through interaction, correlated with the jet. 

The full potential of jets as probes of QGP dynamical properties relies on the ability to have sufficient theoretical control of the totality of modifications imparted on a jet by its interaction with QGP to devise phenomenological frameworks capable of yielding results from which a bona fide comparison with experimental measurements can be made. 
Any \textit{bona fide} extraction of physical information from experimental jet results through comparison with theoretical predictions requires that those theoretical computations, be then purely analytical or the result of Monte Carlo event generation, are subjected to a workflow analogous to the one used to reconstruct experimentally. 

The contribution of QGP response to jets in \textsc{Jewel} was originally addressed in \cite{KunnawalkamElayavalli:2017hxo} where two different background subtraction procedures -- grid subtraction and four-momentum subtraction -- were proposed. Results obtained from grid based subtraction were found to be very dependent on the chosen granularity of the grid. While four-momentum subtraction performed well for all observables computed exclusively from transverse momenta of jet constituents, it led to potentially unphysical results for observables, such as the jet invariant mass, dependent on the full four-momenta of jet constituents.    

In this work we put forward a background subtraction procedure that accounts for the specificity of how QGP response is modelled in the \textsc{Jewel} event generator and which yields robust results for all observables, importantly those computed from the full four-momenta of jet constituents.

The paper is organized as follows. In sec.\ref{sec:jewel} we provide a brief description of \textsc{Jewel} MC with a specific focus on the modelling of QGP response. In sec.\ref{sec:4momsubtraction} we identify the origin of the generation of unphysical negative squared jet masses by four momentum subtraction. Sec.\ref{sec:constituentsub}, introduces a new subtraction method based on the widely used constituent subtraction. Results for typical jet shape and sub-structure observables are discussed in sec.~\ref{sec:results} with particular emphasis on subtraction, and sec.~\ref{sec:conclusions} presents some conclusions.

\section{JEWEL}
\label{sec:jewel}

\textsc{Jewel}~\cite{Zapp:2012ak,Zapp:2013vla} simulates the QCD evolution and re-scattering of hard partons in a background medium. It builds on PYTHIA\,6.4~\cite{Sjostrand:2006za}, which is used for hard matrix elements, initial state parton shower and hadronisation. \textsc{Jewel} implements the final state parton shower interleaved with re-scattering and provides a simple model for the background medium. Re-scattering in the medium is described by LO matrix elements supplemented with a parton shower, such that both elastic and inelastic scattering are included with leading-log correct relative contributions. The interplay between different sources of radiation is governed by the formation time in such a way that always the emission with the shortest formation time gets realised. The coherence between several scattering processes within the formation time of the same emission giving rise to the LPM effect is included in a probabilistic formulation.

\textsc{Jewel} does not simulate complete heavy ion events, thermal background partons that do not participate in a scattering involving a hard parton never explicitly show up in the event. Instead, the event consists of partons coming from the hard matrix elements and the parton showers. It is possible to also keep the thermal partons that did interact with a hard parton (``recoils'') in the event, but then the thermal momentum these partons had before the interaction has to be subtracted from the final event. Two options for doing this, grid subtraction and four-momentum subtraction, are discussed in~\cite{KunnawalkamElayavalli:2017hxo}. Grid based subtraction has not been pursued further due to short-comings of the method. The four-momentum subtraction methods and its limitations will be briefly reviewed in the next section before a new method along the lines of constituent subtraction is introduced in section~\ref{sec:constituentsub}.

\section{Four-momentum subtraction and negative squared jet masses}
\label{sec:4momsubtraction}

Four-momentum subtraction, introduced in \cite{KunnawalkamElayavalli:2017hxo}, removes the thermal momenta exactly from the jet’s four-momentum. The thermal momenta to be subtracted, those that in the absence of the jet would have ended up within the jet's reconstruction region, are determined by adding to the final state particles list a set of `dummy' neutral particles with very small energy and momenta and pointing in the direction of the thermal momenta\footnote{In the original version of the four-momentum subtraction, the dummy particles were massless and had the same azimuthal angle and pseudo-rapidity as the thermal momentum. Since \textsc{Jewel}-\,2.3.0 the dummies are massive and have the same azimuthal angle, pseudo-rapidity, and rapidity as the thermal momentum. This is preferable since the distance measure in the $\kt$-family of jet clustering algorithms is based on rapidity separation. The only observable where this change has any visible consequences is the jet mass, which is particularly sensitive to soft particles at the edge of the jet.}. In operational terms, these `dummy' particles are the same as the ghosts used by FastJet \cite{Cacciari:2011ma} during its clustering to determine the jet area. They can get clustered into jets without affecting the jet’s momentum or structure. Thermal momenta, that are matched to a dummy (in the azimuthal angle - rapidity plane) inside a jet, are subtracted from the jet’s momentum. The resulting four vector constitutes the subtracted jet momentum. The procedure was implemented as follows:
\begin{enumerate}
    \item Cluster the initial jet collection from the final state particles (including dummy particles);
    \item Compile a list of the thermal momenta (particles in the HepMC event record with status code 3);
    \item For each jet, get the list of thermal momenta that are within $\Delta R < 1 \cdot 10^{-5}$ of a the jet constituent, i.e a dummy particle;
    \item Sum up the four-momenta of the matched thermal momenta. This constitutes the background;
    \item For each jet subtract the background four-momentum from the jet’s four momentum, this provides the corrected jet collection.
    \item Calculate jet observables from corrected jet four-momenta.
\end{enumerate}

Four-momentum subtraction can also be performed for suitably defined jet sub-structure observables. It then operates not on the entire jet, but only a part of it, e.g. a sub-jet. In iterative procedures this means that the subtraction also has to be performed iteratively, i.e.\ in every step.

\smallskip

Due to the way in which a parton level quantity, i.e. the thermal momenta, is subtracted from jet level four-momenta, four-momentum subtraction only works for IRC safe quantities. When cuts are applied on the hadron distribution subtraction is in general not possible any more, because it is not known which part of the thermal momenta should be subtracted from the reduced hadron population. This can lead to ambiguities when comparing to measurements.

\medskip

\begin{figure}[h]
    \centering 
    \includegraphics[width=.75\textwidth]{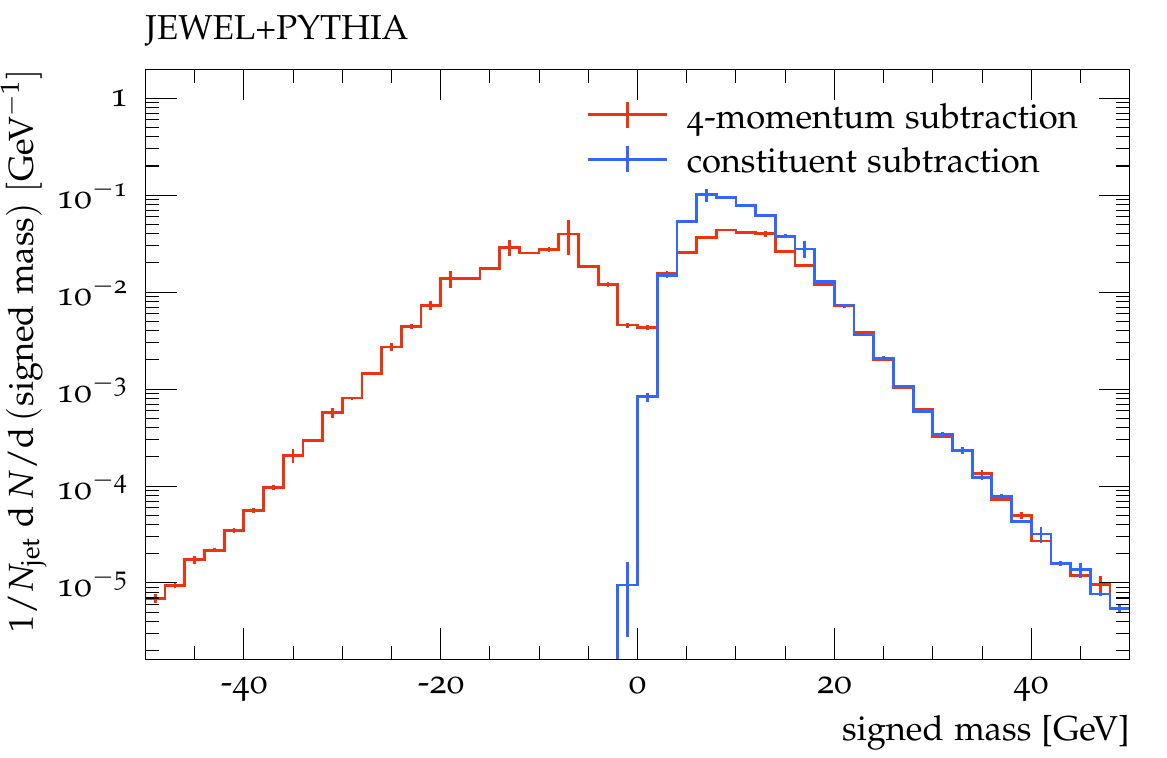}
    \caption{Signed mass, i.e. $\mathrm{sgn}(M_\text{jet}^2)\sqrt{|M_\text{jet}^2|}$, distribution after four-momentum (blue) and constituent subtraction (red) for $R=0.4$ anti-$k_\perp$ jets with $p_\perp^\text{(jet)} > \unit[100]{GeV}$ and $|\eta^\text{(jet)}| < 2.5$ from a di-jet sample at $\sqrt{s_\text{NN}} = \unit[2.76]{TeV}$}
    \label{fig:negmass}
\end{figure}

A problem with this four-momentum subtraction method is that the jet mass, although formally an IRC safe quantity, behaves in a strange way. In particular, the squared jet mass is negative for roughly half of the jets (Fig.~\ref{fig:negmass}). 

To understand how this happens we consider the elastic scattering of a hard parton off a thermal parton $p^\mu_\text{h} + p^\mu_\text{th} = p'^\mu_\text{h} + p^\mu_\text{rec}$. If all momenta are contained inside the reconstructed jet, the jet momentum after four-momentum subtraction can be written as
\begin{equation}
    {P'}^\mu = p^\mu_\text{rem} + {p'}^\mu_\text{h} + p^\mu_\text{rec} - p^\mu_\text{th} = p^\mu_\text{rem} + {p'}^\mu_\text{h} + q^\mu = p^\mu_\text{rem} + p^\mu_\text{h} = P^\mu\,,
\end{equation}
where $q^\mu$ is the four-momentum transfer in the scattering and $p^\mu_\text{rem}$ is the momentum of the remainder system. In this case the jet momentum does not change and the jet mass after scattering is the same as the jet mass before. If, however, one or several of the partons involved in the scattering fall outside the jet cone, the jet mass changes due to the scattering. The different possibilities of losing a parton are listed below.
\begin{enumerate}
    \item The thermal momentum is not contained in the reconstructed jet. Then the jet momentum becomes
    \begin{equation}
     {P'}^\mu = p^\mu_\text{rem} + {p'}^\mu_\text{h} + p^\mu_\text{rec} = p^\mu_\text{rem} + p^\mu_\text{h} + p^\mu_\text{th} = P^\mu  + p^\mu_\text{th}
    \end{equation}
    and for the squared jet mass one gets
    \begin{equation}
        {M'}_\text{jet}^2 = M_{\text{jet}}^2 + m^2_\text{th} + 2 P_\mu p^\mu_\text{th} > {M'}_\text{jet}^2 \,.
    \end{equation}
    The jet mass thus increases slightly, since the scalar product of two physical 4-momenta cannot be negative. Given the typical kinematics ($m_\text{th}$ small, $p^\mu_\text{th}$ a soft momentum, $P_\mu$ a hard momentum with mass small compared to energy) the increase is mostly rather small.
    \item The recoiling thermal parton is outside the jet. Then the jet momentum becomes
    \begin{equation}
     {P'}^\mu = p^\mu_\text{rem} + {p'}^\mu_\text{h} - p^\mu_\text{th} = p^\mu_\text{rem} + p^\mu_\text{h} - p^\mu_\text{rec} = P^\mu  - p^\mu_\text{rec}
    \end{equation}
    and for the squared jet mass one gets
    \begin{equation}
        {M'}_\text{jet}^2 = M_\text{jet}^2 + m^2_\text{rec} - 2 P_\mu p^\mu_\text{rec} \,.
    \end{equation}
    Now the squared jet mass can decrease and even become slightly negative. Again, the scalar product $P_\mu p^\mu_\text{rec}$ is typically small, so there is only a mild change in jet mass. Moreover, it is overcompensated by the first case, which is more likely to occur. 
    \item The scattered hard parton is outside the jet. Then the jet momentum becomes
    \begin{equation}
     {P'}^\mu = p^\mu_\text{rem} + p^\mu_\text{rec} - p^\mu_\text{th} = p^\mu_\text{rem} + q = P^\mu  - p^\mu_\text{h}
    \end{equation}
    and for the squared jet mass one gets
    \begin{equation}
        {M'}_\text{jet}^2 = M_\text{jet}^2 + m^2_\text{h} - 2 P_\mu p^\mu_\text{h} \,.
    \end{equation}
    Here $m^2_\text{h}$ is the virtual mass squared of the hard parton, which is normally small compared to its momentum. But now $P_\mu p^\mu_\text{h}$ is the scalar product of two hard momenta and given typical kinematics it is not small compared to the squared jet mass. This situation can thus lead to a sizeable reduction of the squared jet mass, usually turning it negative. This case is also frequent enough to lead to the large effects observed in figure~\ref{fig:negmass}.
\end{enumerate}
Negative squared jet masses thus appear frequently with four-momentum subtraction, but should be regarded as an artifact rather than a natural feature of the method. The squared mass of any composite object consisting of particles is always positive. If one could reconstruct a jet in a heavy ion environment and then look at its constituents one by one and keep only those that receive most of their energy from hard partons rather than the background, then the squared jet mass could never be negative. Since this is impossible due to the quantum mechanical nature of particle collisions, one has to resort to other background subtraction techniques both in experiment and theory. Since it is desirable that the background subtraction leaves the squared jet mass positive (and doesn't introduce other artifacts in the jet mass distribution), we switch to constituent subtraction, which we discuss in the next section.

\section{Background subtraction with constituent subtraction}
\label{sec:constituentsub}

The new background subtraction procedure\footnote{an implementation of the procedure as a \textsc{Rivet} projection is publicly available from  
\href{http://jewel.hepforge.org}{jewel.hepforge.org}.} uses the constituent subtraction method~\cite{Berta:2014eza} with the ghosts replaced by the thermal momenta. In contrast to the old four-momentum subtraction the thermal four-momentum is not subtracted locally, but its transverse momentum and mass are subtracted from nearby particles in a way that ensures positive definite mass squares (cf.\,Fig.~\ref{fig:negmass}). The exact algorithm is repeated below.

All four-momenta (final state particles and thermal momenta) are represented by their transverse momentum $p_\perp$, mass $m_\delta = \sqrt{m^2 + p_\perp^2} - p_\perp$, rapidity $y$ and azimuthal angle $\phi$:
\[ p^\mu = \left( (m_\delta + p_\perp) \cosh(y),\ p_\perp \cos(\phi),\ p_\perp \sin(\phi),\ (m_\delta + p_\perp) \sinh(y) \right) \]
A list of all possible pairs consisting of a final state particle $i$ and a thermal momentum $k$ is compiled and sorted by distance, where the distance measure used is
\[ \Delta R_{ik} = \sqrt{(y_i-y_k)^2 + (\phi_i-\phi_k)^2} \ . \]
The subtraction proceeds by going through the list (starting from the smallest distance) and in each pair subtracting the smaller $p_\perp$ from the larger and the smaller $m_\delta$ from the larger:
\begin{equation}
\begin{array}{lcl}
\text{if} \quad p_{\perp}^{(i)} \ge p_{\perp}^{(k)} & \quad : \qquad & p_{\perp}^{(i)} \to p_{\perp}^{(i)} - p_{\perp}^{(k)} \\
                                 &  & p_{\perp}^{(k)} \to 0\\
\text{if} \quad p_{\perp}^{(i)} < p_{\perp}^{(k)} & \quad : \qquad & p_{\perp}^{(i)} \to 0\\
                                 &  & p_{\perp}^{(k)} \to p_{\perp}^{(k)} - p_{\perp}^{(i)}
\end{array}
\end{equation}
and
\begin{equation}
\begin{array}{lcl}
\text{if} \quad m_\delta^{(i)} \ge m_\delta^{(k)}& \quad : \qquad & m_\delta^{(i)} \to m_\delta^{(i)} - m_\delta^{(k)} \\
                                 &  & m_\delta^{(k)} \to 0\\
\text{if} \quad m_\delta^{(i)} < m_\delta^{(k)}& \quad : \qquad & m_\delta^{(i)} \to 0\\
                                 &  & m_\delta^{(k)} \to m_\delta^{(k)} - m_\delta^{(i)}
\end{array}
\end{equation}
One can continue until the end of the list is reached, or stop at some distance cut-off in order to avoid subtractions from particles that are far away. In our implementation we chose the latter option with a cut-off at 0.5 in $\Delta R_{ik}$. After all subtractions are done, all momenta with $p_\perp = 0$ are removed. The remaining momenta constitute the subtracted ensemble.

Inside jets the final state particles carry much more momentum than the thermal momenta. The constituent subtraction procedure thus leads to all thermal momenta being used up and disappearing from the ensemble. Since only final state particles survive, one can keep track of the particles' flavour (and other properties if needed) and arrive after subtraction at a final state consisting of identified particles. One can then extract observables such as charged jet mass or jet fragmentation functions that are built from certain particle species only (in this example charged particles). This is not possible with the old four-momentum subtraction and -- at least for the charged jet mass -- improves the agreement with data, as discussed in section~\ref{subsec:jetmass}.

\subsection{Subtraction on reconstructed jets}
\label{subsec:jetlevel}

Constituent background subtraction can be carried out on all final state particles prior to jet reconstruction, or after jet reconstruction on the particles clustered into jets. In the latter version it can be directly compared to four-momentum subtraction, which always operates on reconstructed jets.

When subtracting on jet level, it first has to be determined which of the thermal momenta should be subtracted. To this end dummy particles, which have very small momenta and the same rapidity, pseudo-rapidity and azimuth as the thermal momenta, are added to the final state. After jet reconstruction one then finds the dummy particles that got clustered into the jet, the corresponding thermal momenta should then be subtracted.

\begin{figure}
\centering
\includegraphics[width=.45\textwidth]{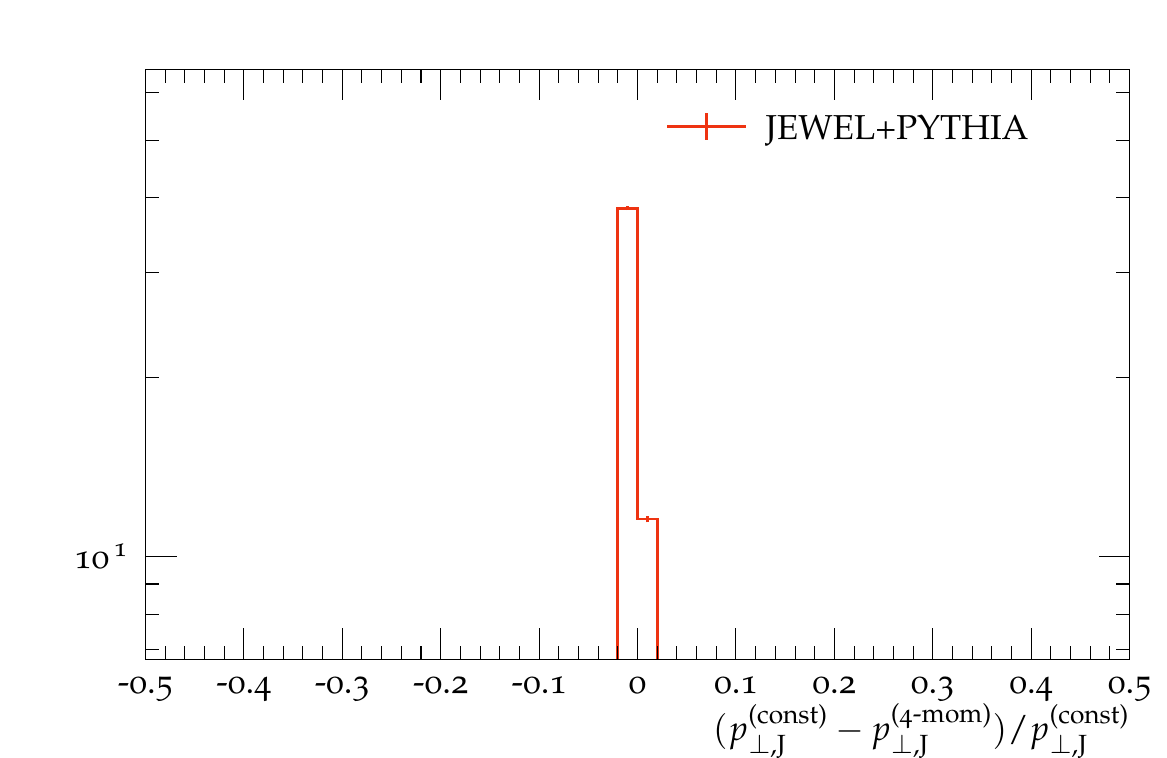}
\includegraphics[width=.45\textwidth]{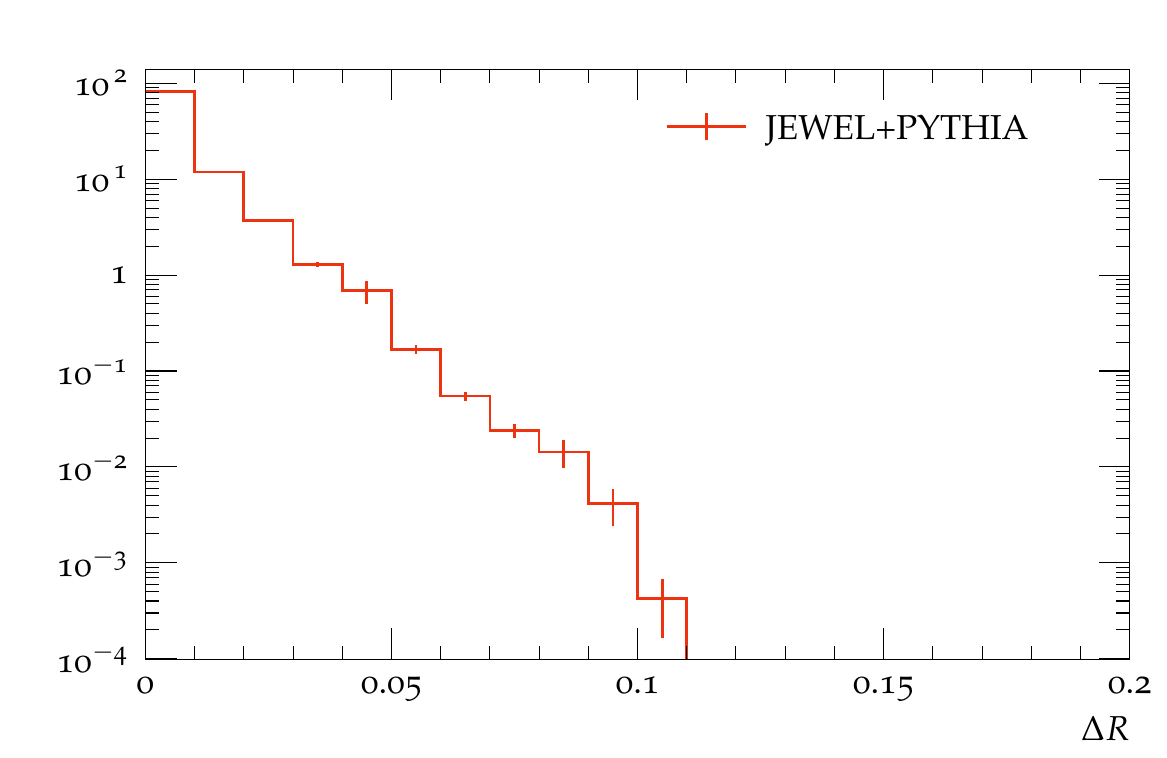}\\
\includegraphics[width=.45\textwidth]{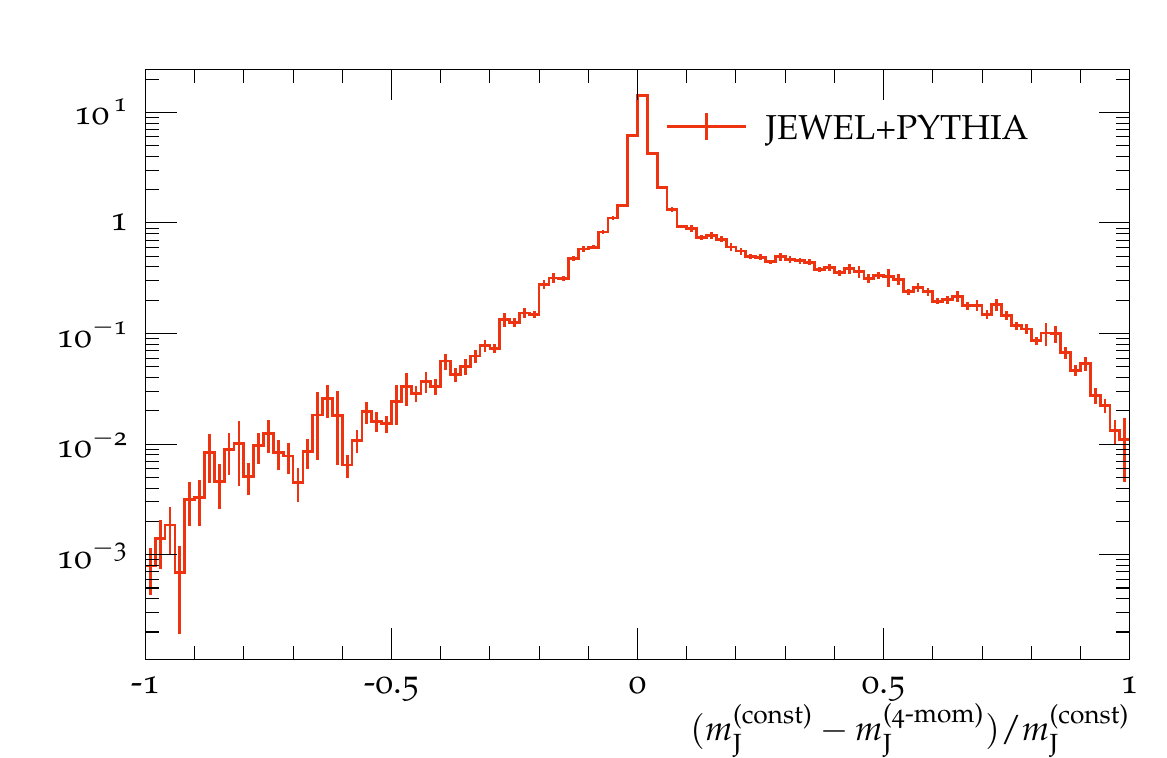}
\caption{Comparing four-momentum and constituent background subtraction for $R=0.4$ anti-$k_\perp$ jets from a di-jet sample at $\sqrt{s_\text{NN}} = \unit[2.76]{TeV}$ with $p_\perp^\text{(jet)} > \unit[100]{GeV}$ and $|\eta^\text{(jet)}| < 2.5$: relative difference in jet transverse momentum (top left), angle $\Delta R = \sqrt{(\Delta \phi)^2 + (\Delta \eta)^2}$ between jet axes (top right), and relative mass difference (bottom). The mass difference is shown only for jets where the squared mass with four-momentum subtraction is positive.}
\label{fig:4mom-const-comp}
\end{figure}

In Fig.~\ref{fig:4mom-const-comp} four-momentum and constituent background subtraction are compared. To this end a di-jet sample is generated with \textsc{Jewel}\,2.3.0~\cite{Zapp:2012ak,Zapp:2013vla} in the standard set-up using the simple medium model described in~\cite{Zapp:2013vla}. Jets are reconstructed with the anti-$k_\perp$ algorithm~\cite{Cacciari:2008gp} and for each the background subtraction is performed twice: once with the four-momentum method and once with constituent subtraction. As seen in the left panel of Fig.~\ref{fig:4mom-const-comp} the jet's transverse momentum after background subtraction is the same with both methods while the jet axis can move to a slightly different position (central panel in Fig.~\ref{fig:4mom-const-comp}). The mass difference, however, is clearly asymmetric (right panel in Fig.~\ref{fig:4mom-const-comp}): the mass with constituent background subtraction is smaller than with four-momentum subtraction (considering only jets with a positive mass squared after four-momentum subtraction). Observables built from the jet's transverse momentum and direction (e.g. the di-jet asymmetry $A_J$) are thus insensitive to details of the background subtraction procedure, while the jet mass is not.

\subsection{Event level subtraction}
\label{subsec:eventlevel}

\begin{figure}
\centering
\includegraphics[width=.45\textwidth]{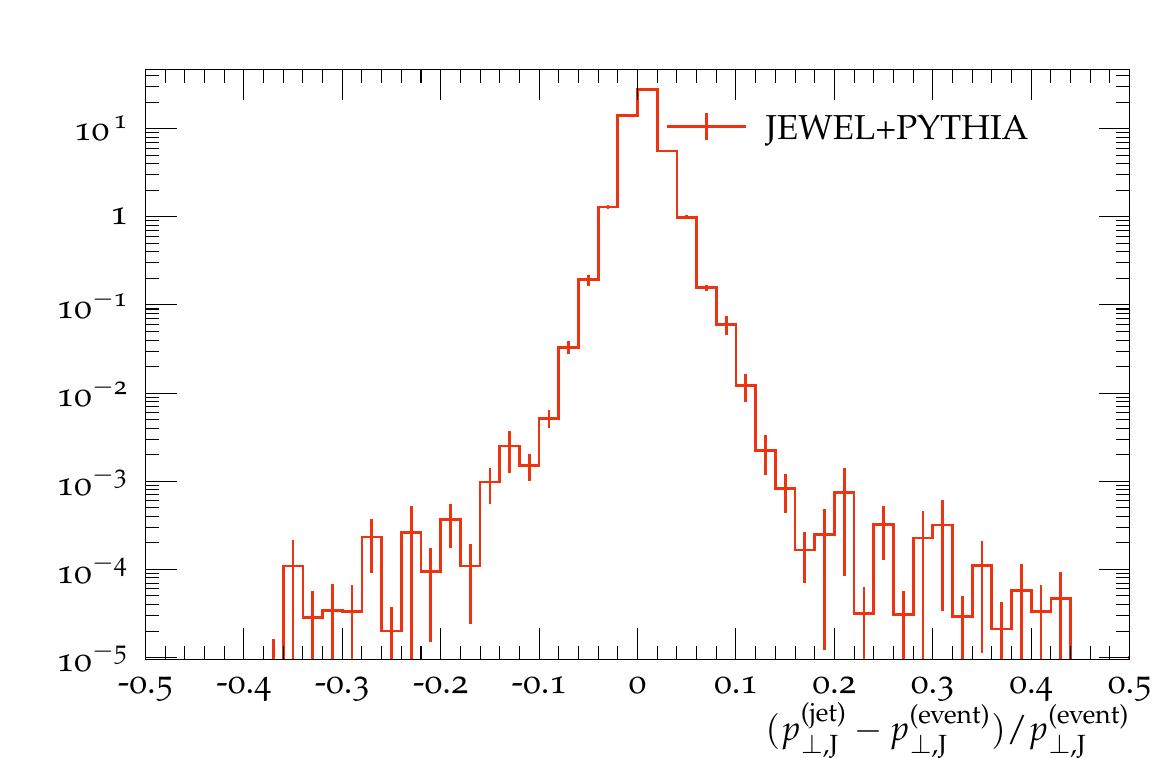}
\includegraphics[width=.45\textwidth]{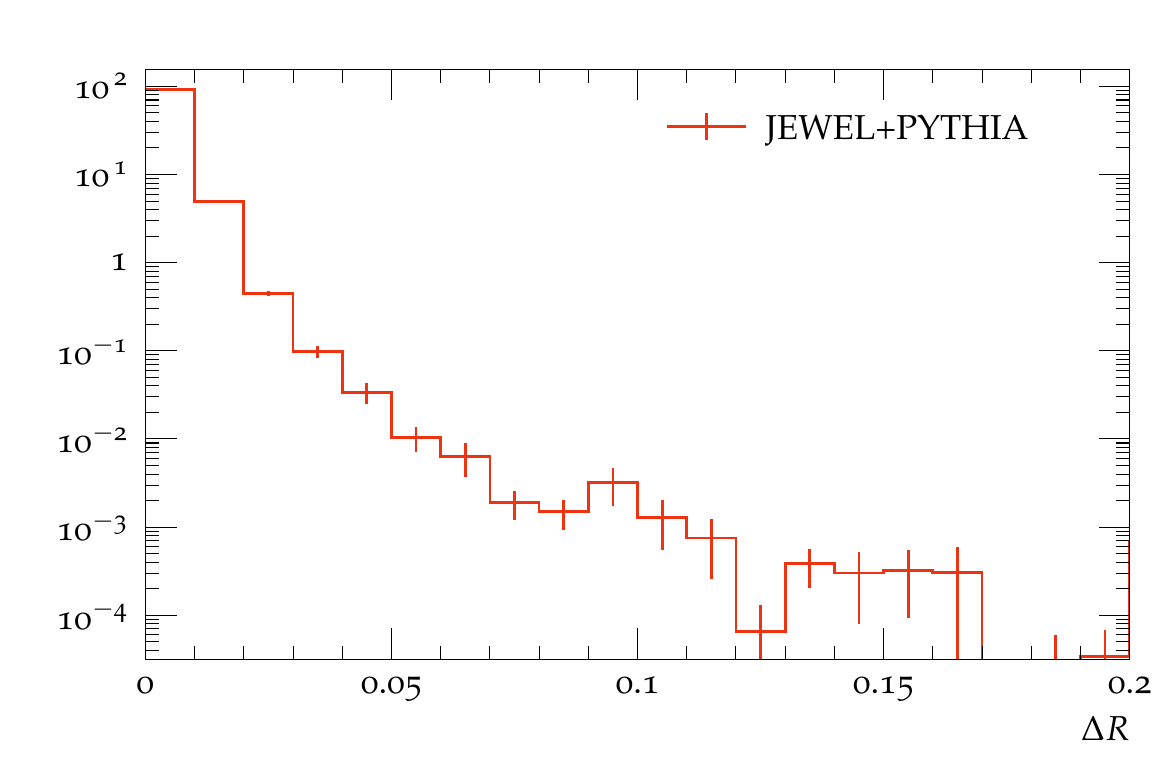}\\
\includegraphics[width=.45\textwidth]{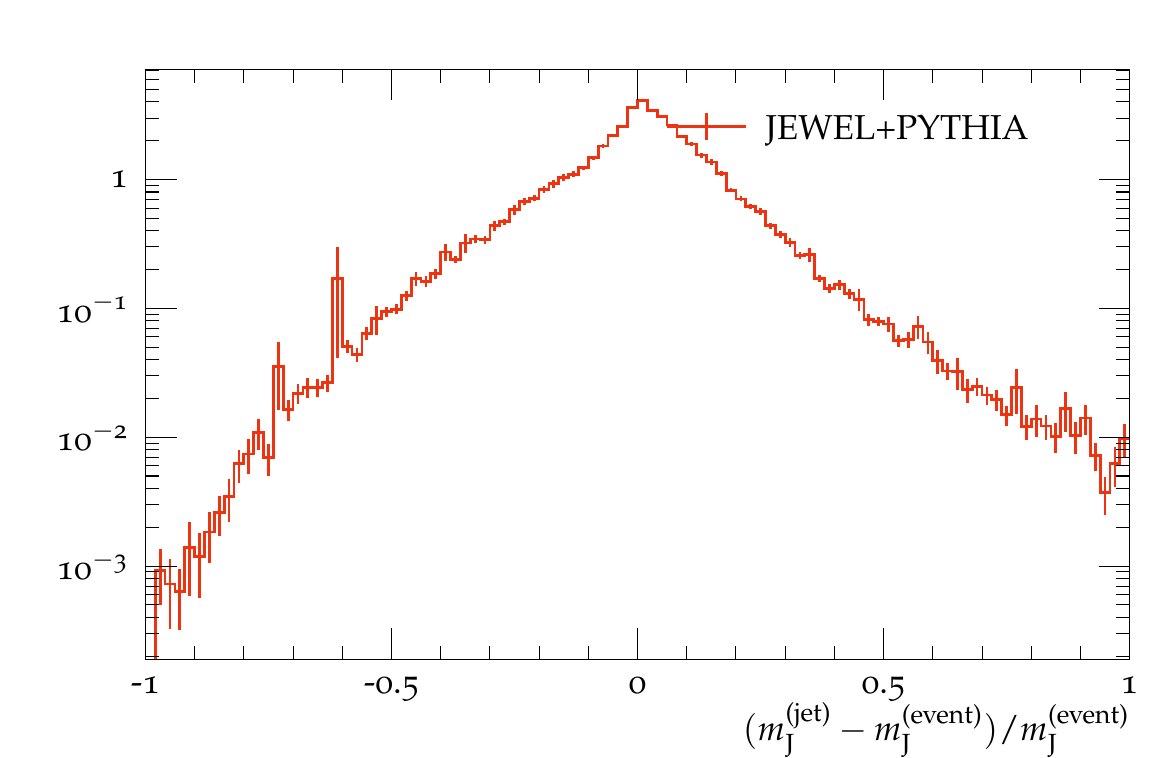}
\caption{Comparing event-wise and jet-wise subtraction for constituent background subtraction for the leading $R=0.4$ anti-$k_\perp$ jet from a di-jet sample at $\sqrt{s_\text{NN}} = \unit[2.76]{TeV}$ with $p_\perp^\text{(jet)} > \unit[100]{GeV}$ and $|\eta^\text{(jet)}| < 2.5$: relative difference in jet transverse momentum (top left), angle $\Delta R = \sqrt{(\Delta \phi)^2 + (\Delta \eta)^2}$ between jet axes (top right), and relative mass difference (bottom).}
\label{fig:const-comp}
\end{figure}

With the constituent subtraction method it is also possible to perform the subtraction of thermal momenta on the entire event before jets are reconstructed. This has the advantage that the jets are less biased by fluctuations in the contribution from medium response. Fig.~\ref{fig:const-comp} shows a comparison between event-wise and jet-wise subtraction. To avoid ambiguities when matching jets from both samples we consider only the leading jet in each event. The difference in jet transverse momentum and the distance between the jet axes are narrow distributions centered at zero. The difference in jet mass is considerably wider, but also centered at zero. It has a slight asymmetry that leads to a slightly wider mass distribution for the jet-wise subtraction compared to event-wise subtraction.   

\section{Results}
\label{sec:results}

The results presented in this section were obtained from a standard di-jet sample in Pb+Pb collisions generated with \textsc{Jewel}-2.3.0\footnote{code publicly available from \href{http://jewel.hepforge.org}{jewel.hepforge.org}} using the proton PDF set  \textsc{Cteq6LL}~\cite{Pumplin:2002vw} and the \textsc{Eps09}~\cite{Eskola:2009uj} nuclear PDF set, both provided by \textsc{Lhapdf}~\cite{Whalley:2005nh}. The QGP is modelled with the simple medium model (with parameters  $T_\text{i}=\unit[485]{MeV}$ and $\tau_\text{i}=\unit[0.6]{fm}$ for $\sqrt{s_\text{NN}} = \unit[2.76]{TeV}$~\cite{Shen:2012vn}, and $T_\mathrm{i}=\unit[590]{MeV}$ and $\tau_\mathrm{i}=\unit[0.4]{fm}$~\cite{Shen:2014vra} for $\sqrt{s_\mathrm{NN}} = \unit[5]{TeV}$). The \textsc{Jewel} events were analysed using the \textsc{Rivet} framework~\cite{Bierlich:2019rhm}. Jets were reconstructed using algorithms provided by \textsc{FastJet}~\cite{Cacciari:2011ma}.

\subsection{Jet mass}
\label{subsec:jetmass}

\begin{figure}
\centering
\includegraphics[width=.45\textwidth]{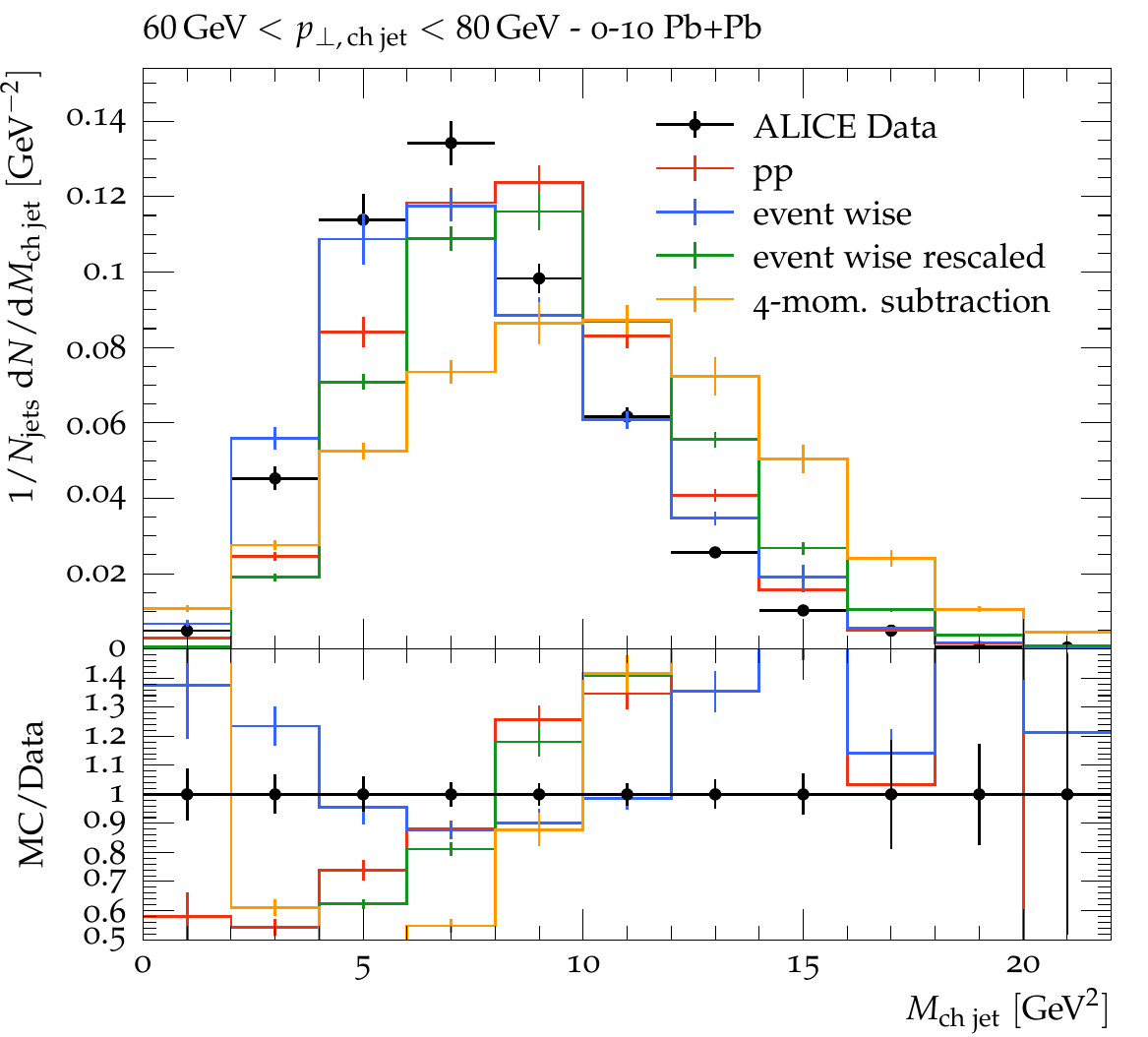}
\includegraphics[width=.45\textwidth]{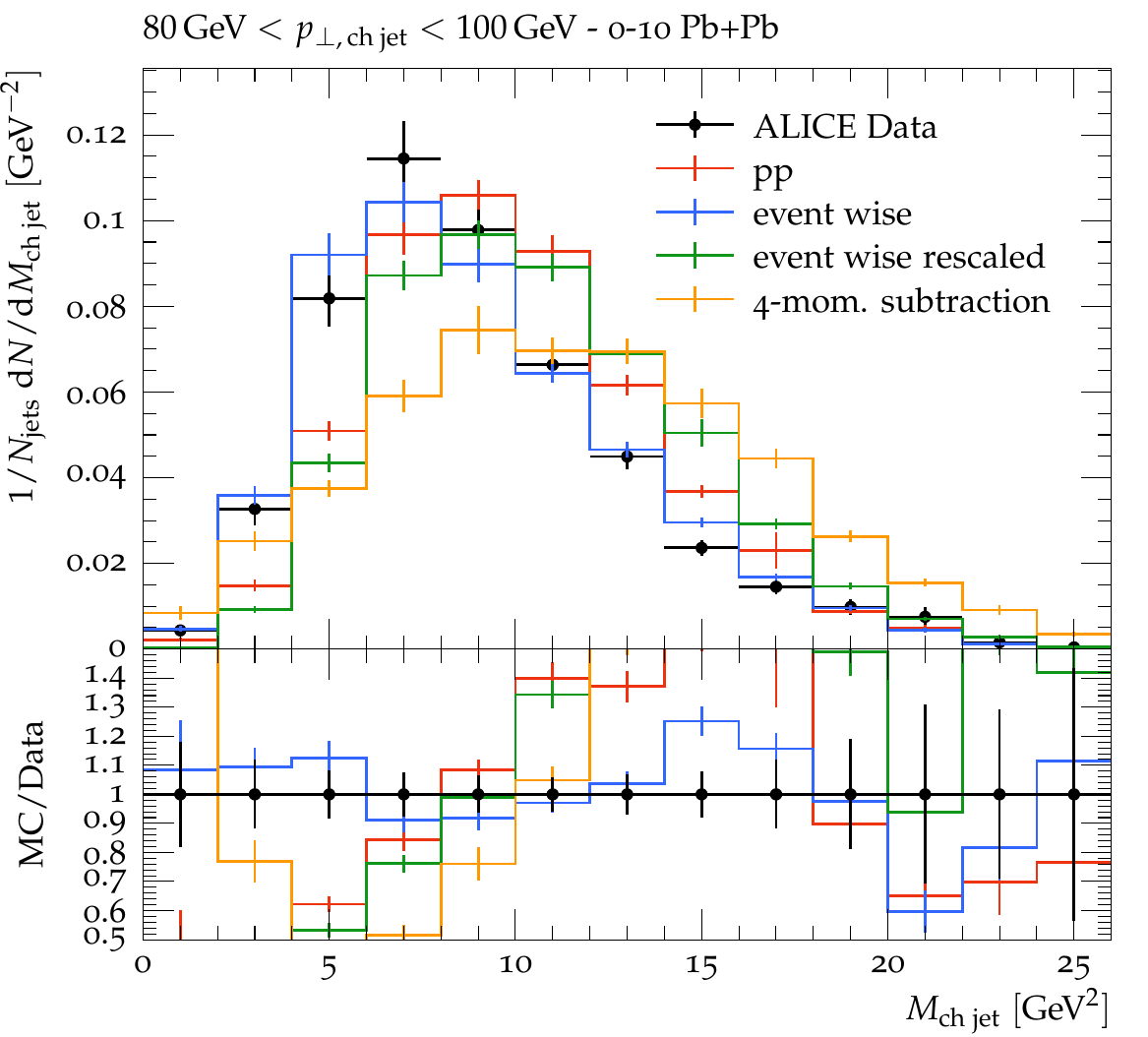}\\
\includegraphics[width=.45\textwidth]{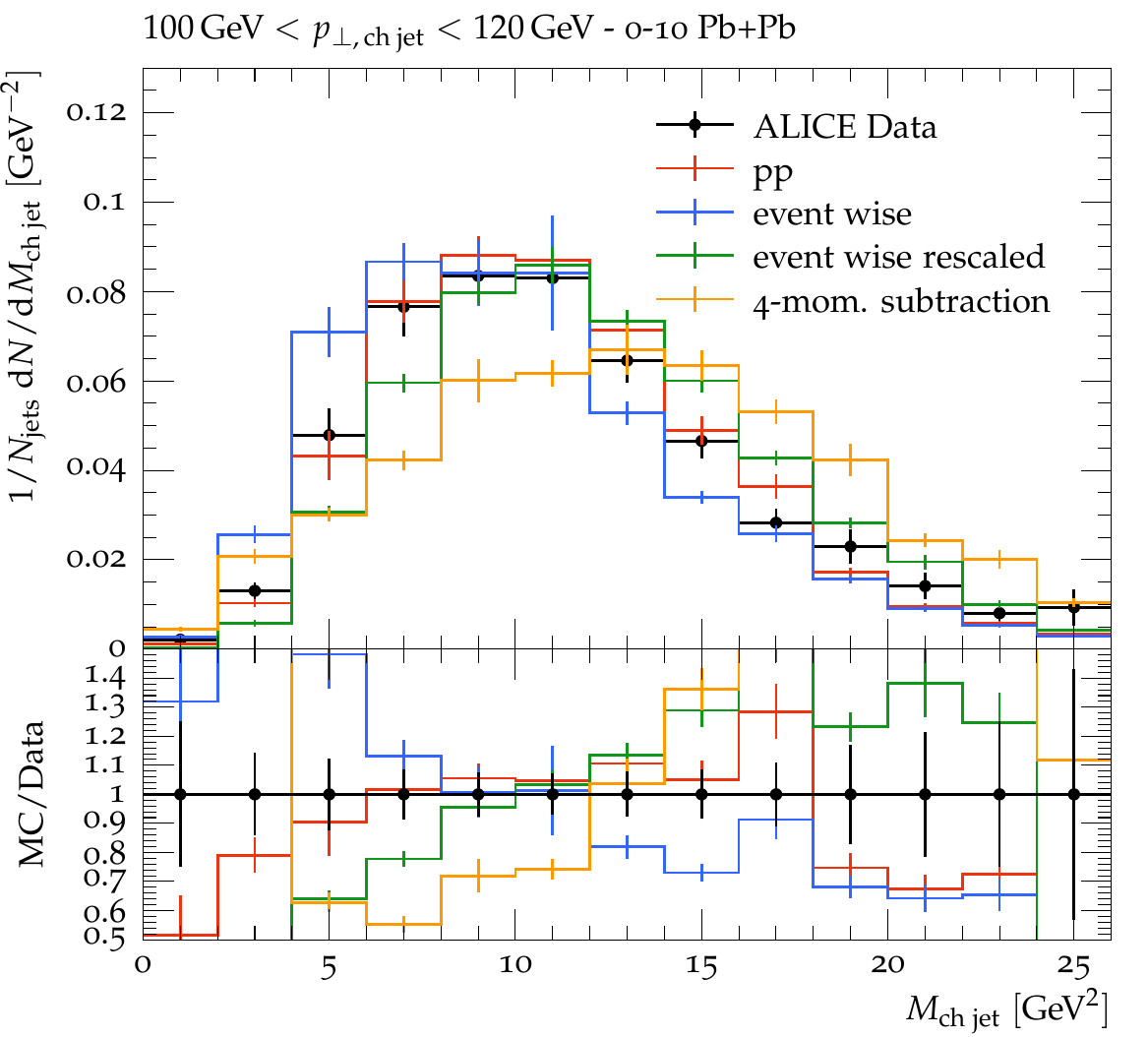}
\caption{Charged jet mass measured by ALICE~\cite{Acharya:2017goa} ($R=0.4$, $|\eta^\text{(jet)}| < 0.5$) with different versions of constituent and four-momentum subtraction. See text for explanations.}
\label{fig:alicejetmass}
\end{figure}

In~\cite{Acharya:2017goa,KunnawalkamElayavalli:2017hxo} it was found that in \textsc{Jewel} with four-momentum subtraction the jet mass was significantly larger than observed in data. Following the discussion in section~\ref{sec:4momsubtraction} we conclude that four-momentum subtraction yields problematic results for jet mass and expect constituent subtraction to be much more reliable. 
ALICE measured the charged jet mass in three different jet $\pt$ bins~\cite{Acharya:2017goa}. With four-momentum subtraction charged jets pose a problem, since it subtracts parton level momenta from hadron level jets. The workaround was to reconstruct and subtract full jets, and afterwards rescale the jet $\pt$ and mass with factors derived from jets with the same cuts in p+p collisions. The results (including only jets with positive squared mass) are shown as the orange histograms in figure~\ref{fig:alicejetmass} and clearly don't agree very well with the ALICE data. Using event wise constituent subtraction with the same procedure of rescaling full jets (green histograms in figure~\ref{fig:alicejetmass}) leads to an improvement, but still not a satisfactory description of the data. The best agreement with data is obtained by doing event wise constituent subtraction preserving the particles' flavour, such that charged jets can be reconstructed in the background subtracted event (blue histograms in figure~\ref{fig:alicejetmass}). The jet mass distribution calculated in this way is shifted slightly towards smaller masses compared to the p+p distribution. This is due to a partial cancellation of the jet collimation effect and medium response. Jet collimation leads to a reduction of the mass (a small mass is correlated to a narrow and hard fragmentation pattern), while medium response increases the mass by adding soft particles distributed widely over the jet (and beyond).

\subsection{Groomed jet mass}

Grooming techniques are used to remove contamination from soft and/or large angle radiation often stemming from background. Several algorithms have been proposed, the most widely used in heavy ion physics is Soft Drop~\cite{Larkoski:2014wba,Dasgupta:2013ihk}. According to this procedure the constituents of jets are first re-clustered with the Cambridge/Aachen algorithm. In an iterative procedure the last re-clustering step is undone to yield two sub-jets. The algorithm terminates when their momenta satisfy the relation
\begin{equation}
    z_g = \frac{\min(p_{\perp,1},p_{\perp,2})}{p_{\perp,1} + p_{\perp,2}} < z_\text{cut} \left(\frac{\Delta R_{12}}{R}\right)^\beta \,,
\end{equation}
where $R$ is the jet radius used in the re-clustering  and $\Delta R_{12}$ the distance between the sub-jets in the rapidity-azimuth plane. Otherwise, the softer sub-jet is discarded and the procedure repeated for the harder one.

\smallskip

In contrast to the jet mass, which is calculated from the four-momenta of all constituents, the groomed jet mass is calculated from the energies and opening angle of the two sub-jets identified by the SoftDrop procedure:
\begin{equation}
\label{eq::mg}
    M_g^2 = 2 E_1 E_2 (1-\cos \theta_{12})
\end{equation}
It is therefore expected to be less sensitive to soft and large angle fragments, and consequently to the subtraction procedure. To demonstrate that this is indeed the case we repeat the analysis by the CMS collaboration~\cite{CMS:2018fof}. Unfortunately, we cannot compare our results directly to the data as the latter are not unfolded and the information on how to smear the Monte Carlo events is not available. 

\begin{figure}
\centering
\includegraphics[width=.45\textwidth]{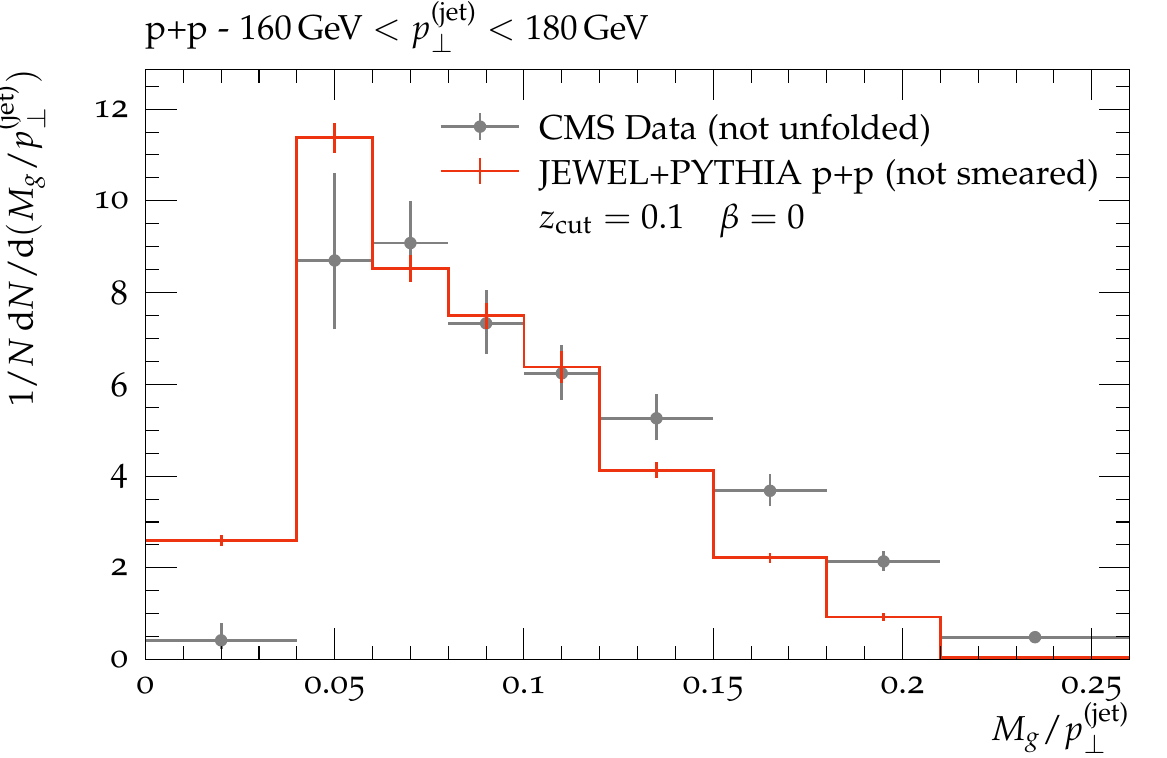}
\includegraphics[width=.45\textwidth]{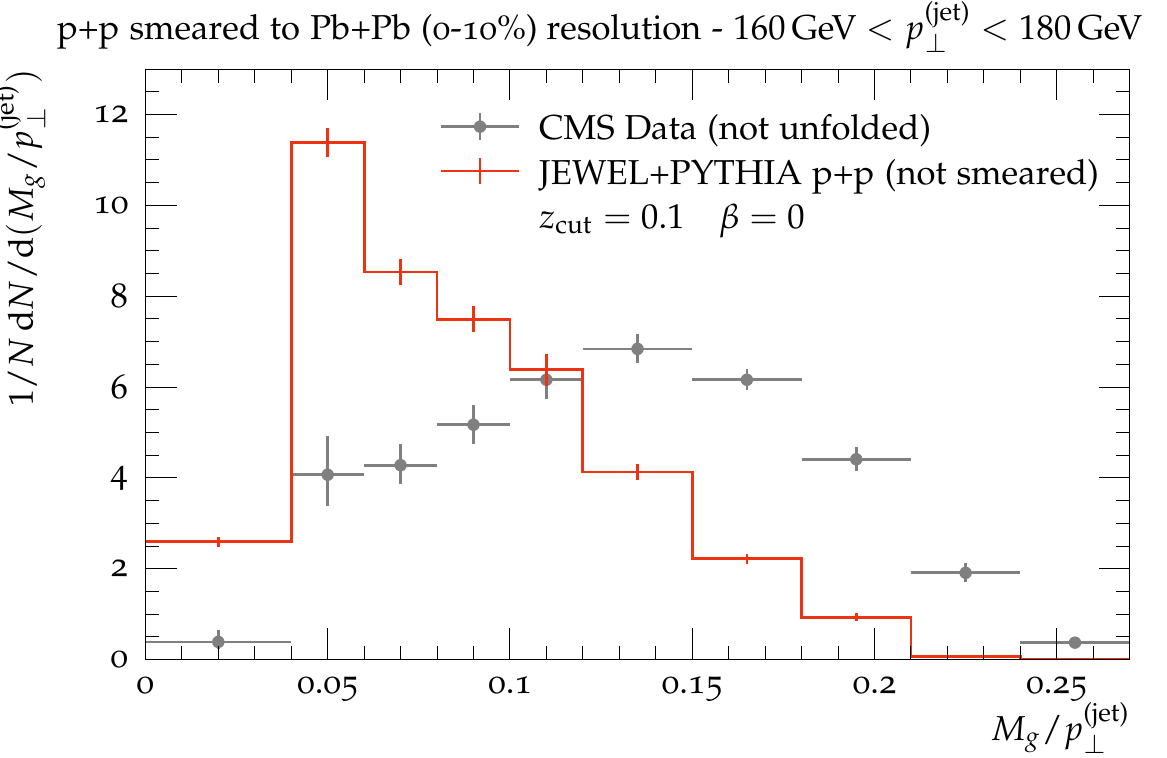}\\
\includegraphics[width=.45\textwidth]{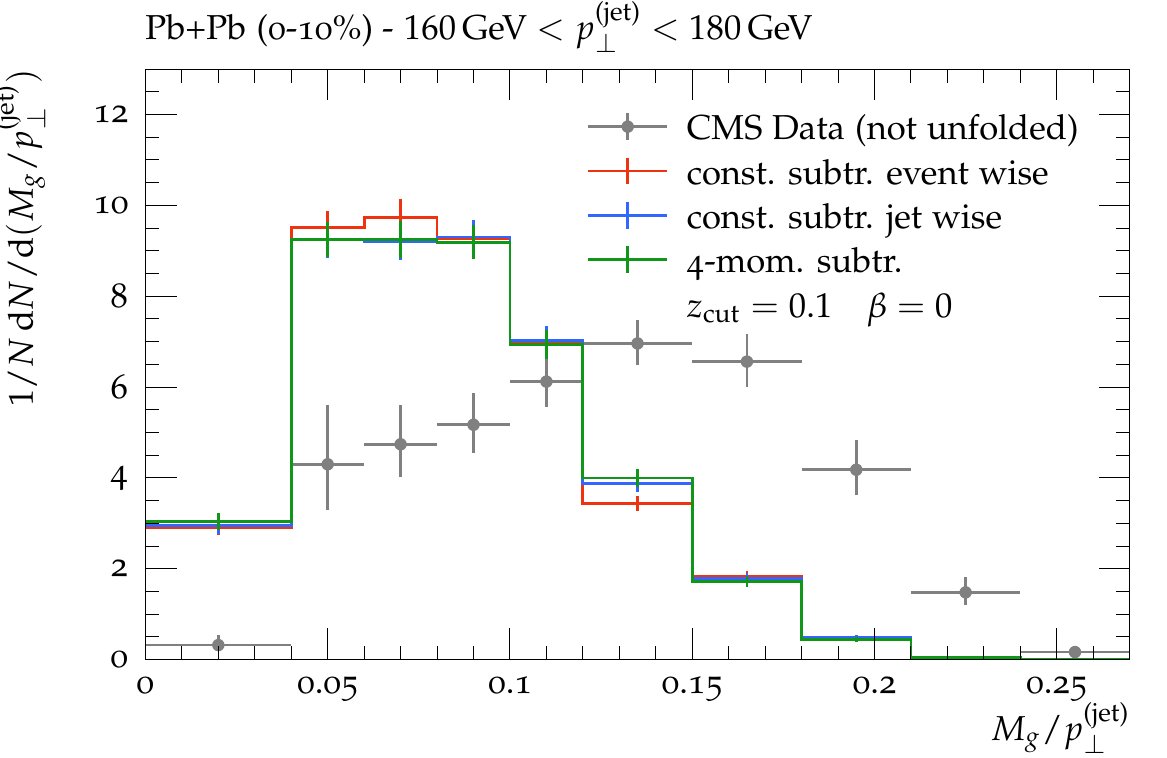}
\includegraphics[width=.45\textwidth]{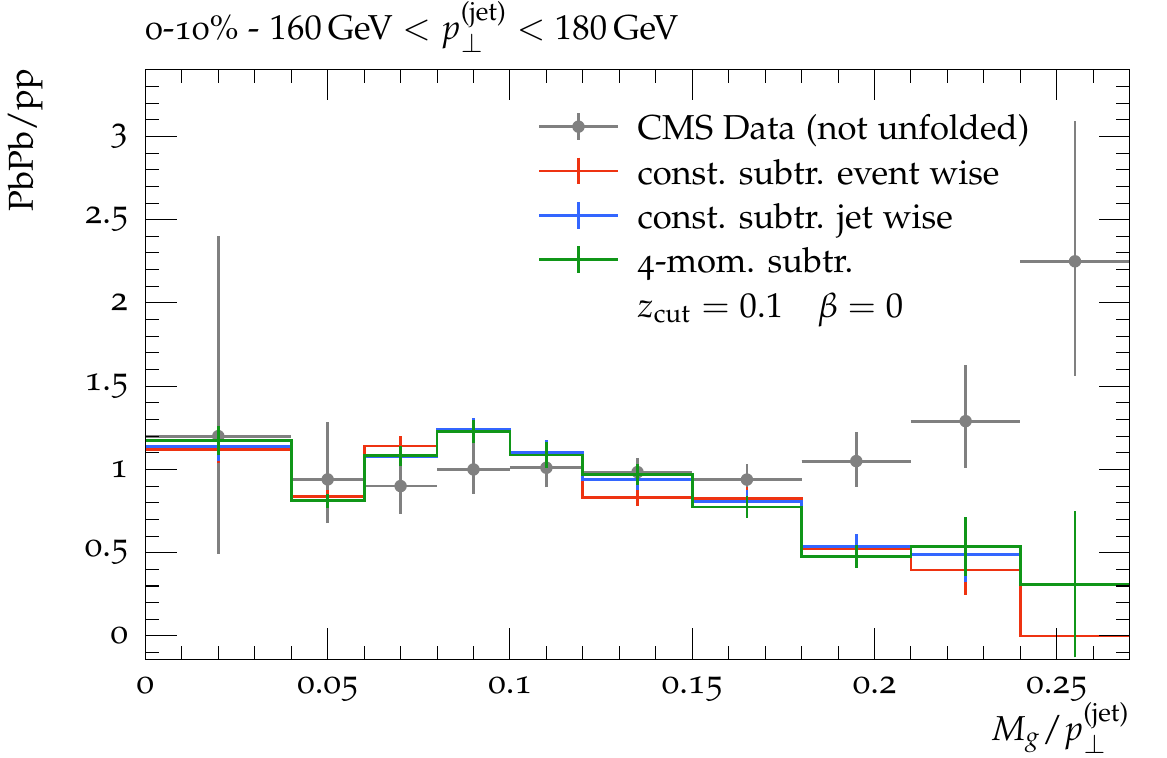}\\
\caption{Groomed jet mass divided by (un-groomed) jet $\pt$ distributions for p+p (top left), p+p smeared to \unit[0-10]{\%} Pb+Pb resolution (top right), \unit[0-10]{\%} Pb+Pb (bottom left) and Pb+Pb (\unit[0-10]{\%}) divided by smeared p+p (bottom right). The SoftDrop parameters are $z_\text{cut} = 0.1$ and $\beta = 0$ and the jets are anti-$\kt$ jets with $R=0.4$,  $|\eta^\text{jet)}| < 1.6$ and $\unit[160]{GeV} < \pt^\text{(jet)} < \unit[180]{GeV}$. CMS data~\cite{CMS:2018fof} and JEWEL+PYTHIA results with different subtraction methods. The CMS measurement has not been unfolded and the JEWEL+PYTHIA results have not been smeared, so a direct comparison is not possible.}
\label{fig:cmsmg1}
\end{figure}

\begin{figure}
\centering
\includegraphics[width=.45\textwidth]{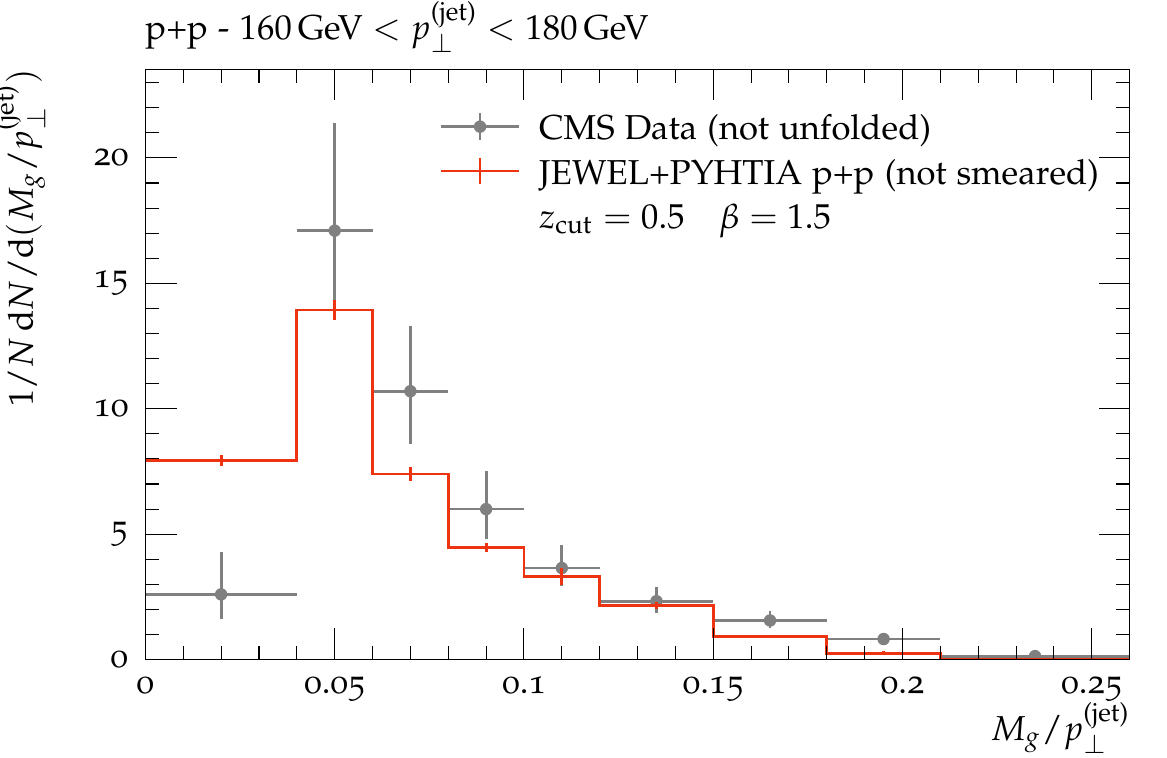}
\includegraphics[width=.45\textwidth]{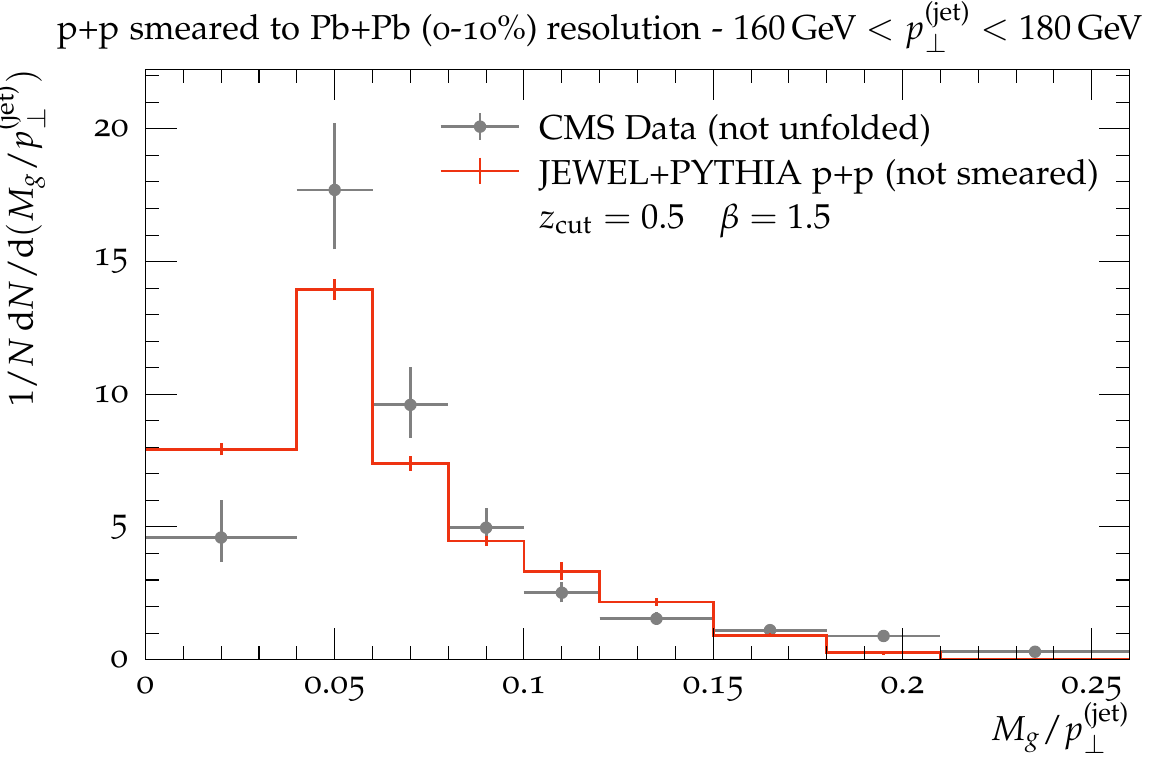}\\
\includegraphics[width=.45\textwidth]{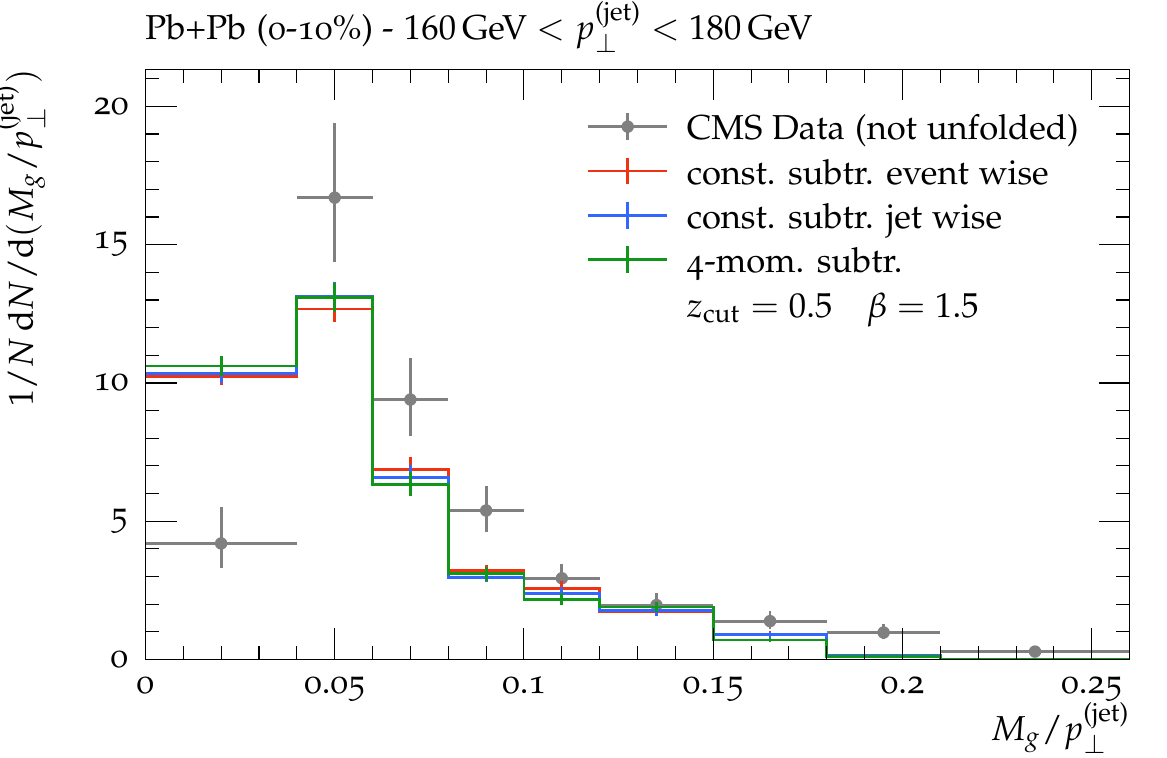}
\includegraphics[width=.45\textwidth]{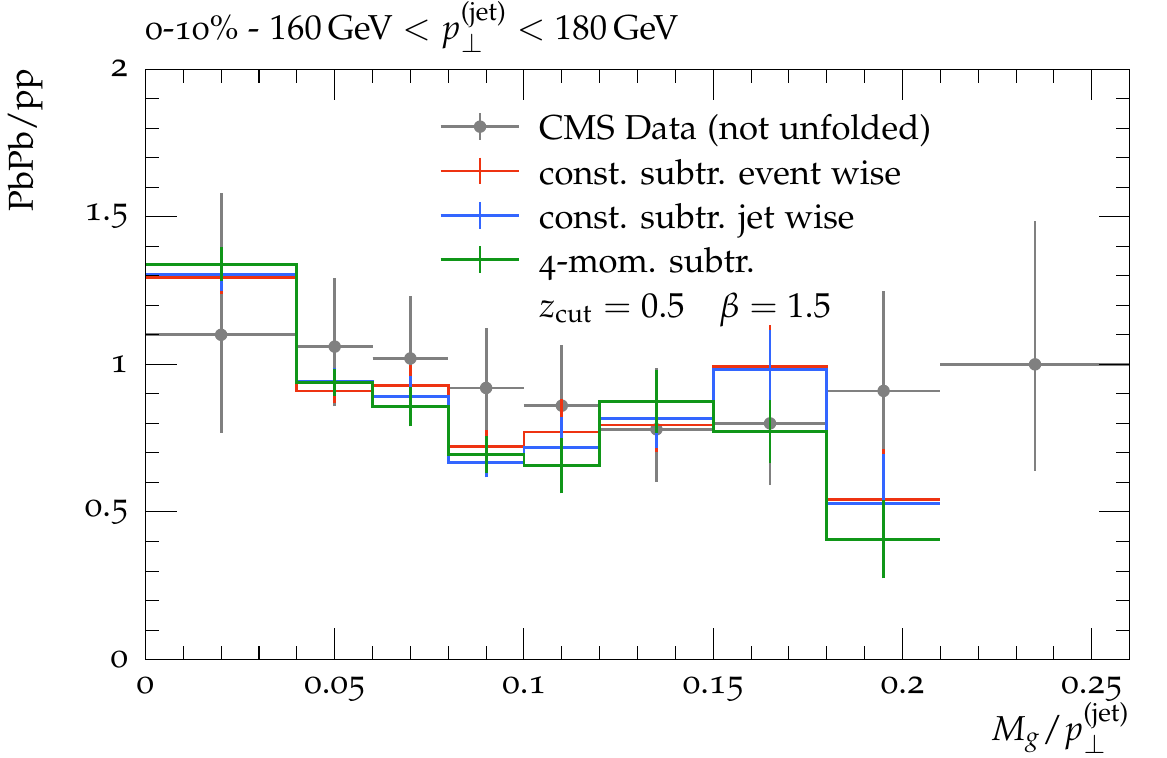}\\
\caption{Groomed jet mass divided by (un-groomed) jet $\pt$ distributions for p+p (top left), p+p smeared to \unit[0-10]{\%} Pb+Pb resolution (top right), \unit[0-10]{\%} Pb+Pb (bottom left) and Pb+Pb (\unit[0-10]{\%}) divided by smeared p+p (bottom right). The SoftDrop parameters are $z_\text{cut} = 0.5$ and $\beta = 1.5$ and the jets are anti-$\kt$ jets with $R=0.4$, $|\eta^\text{jet)}| < 1.6$ and $\unit[160]{GeV} < \pt^\text{(jet)} < \unit[180]{GeV}$. CMS data~\cite{CMS:2018fof} and JEWEL+PYTHIA results with different subtraction methods. The CMS measurement has not been unfolded and the JEWEL+PYTHIA results have not been smeared, so a direct comparison is not possible.}
\label{fig:cmsmg2}
\end{figure}

Figures~\ref{fig:cmsmg1} and \ref{fig:cmsmg2} show the results of the JEWEL+PYTHIA simulation for the ratio of the groomed jet mass (calculated according to equation~\ref{eq::mg}) and the un-groomed jet transverse momentum with two different grooming settings for event wise constituent subtraction, jet wise constituent subtraction and four-momentum subtraction. As expected, there are no significant differences between the three methods, as the groomed jet mass is calculated from the sub-jets' transverse momentum and the opening angle (i.e.\ neglecting the sub-jet invariant masses). A direct comparison to the experimental data is unfortunately not possible. The only conclusion that can be drawn is that there is qualitative agreement with the (un-smeared) p+p data and for the more aggressive grooming setting ($z_\text{cut} = 0.5$ and $\beta = 1.5$), where the smearing effects are not large, also with the Pb+Pb data.

\subsection{Jet - hadron correlations}

The jet-hadron correlations 
\begin{equation}
	\rho(r) = \frac{1}{\pt^\mathrm{(jet)}} \hspace{-2mm} \sum_{\substack{k \mathrm{\ with\ }\\ \Delta R_{kJ} \in [r, r+\delta r]}} \hspace{-3mm} \pt^{(k)} \,,
\end{equation}
studied in this section are a generalisation of the jet profile with the sum running over any set of particles (for the jet profile, it would run only over the constituents of the jet) and $\Delta R_{kJ} = \sqrt{\Delta \phi_{kJ}^2 + \Delta y_{kJ}^2}$ is the angular separation between particle $k$ and the jet axis. When the sum is over all particles the observable is IRC safe and unproblematic for subtraction. However, a typical experimental procedure is to include charged particles above a certain $\pt$. This cannot be reproduced exactly with four-momentum subtraction, which has to include all particles in order to work. With constituent subtraction, however, the experimental procedure can be followed more closely. 

\begin{figure}
\centering
\includegraphics[width=.45\textwidth]{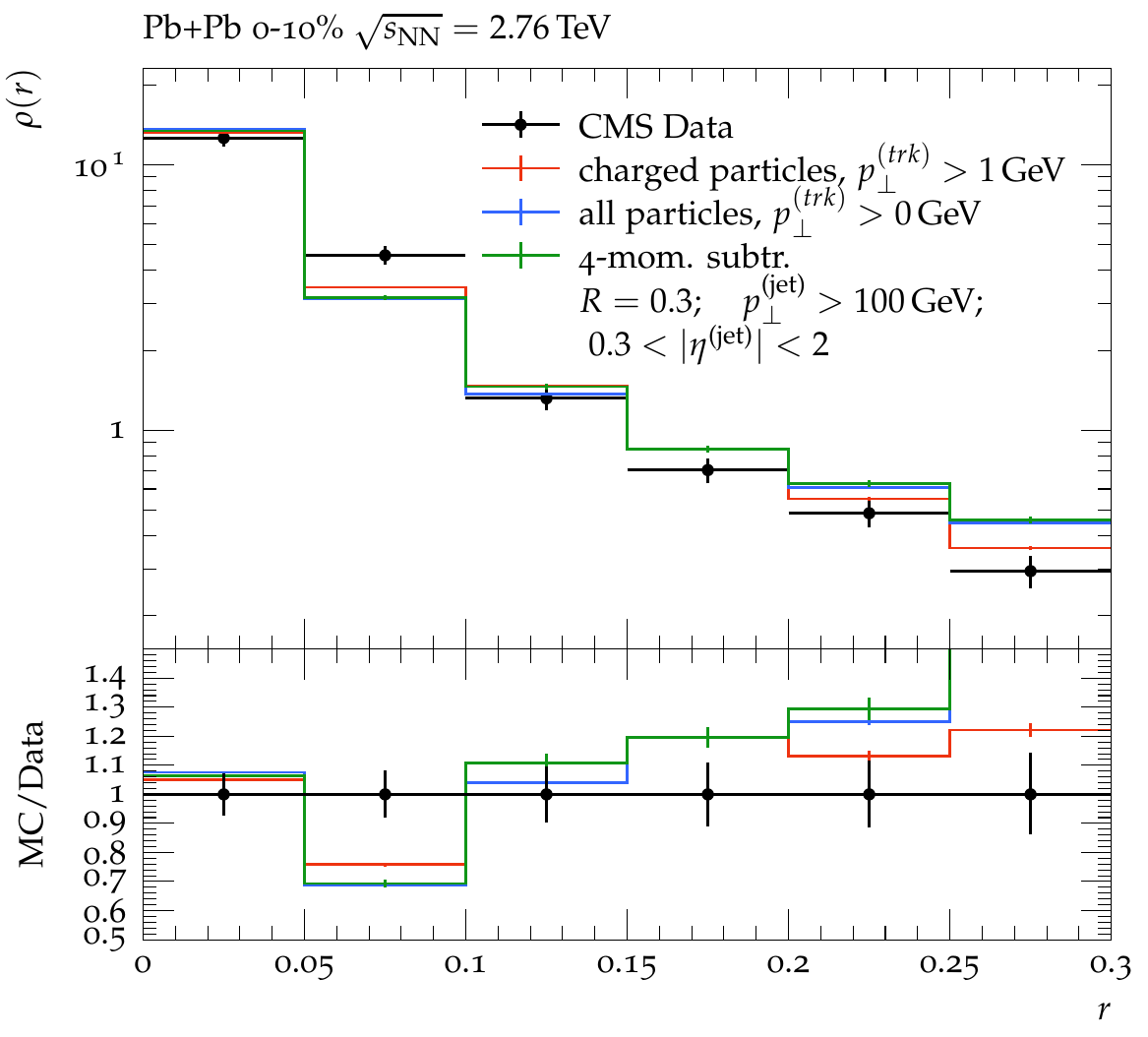}
\includegraphics[width=.45\textwidth]{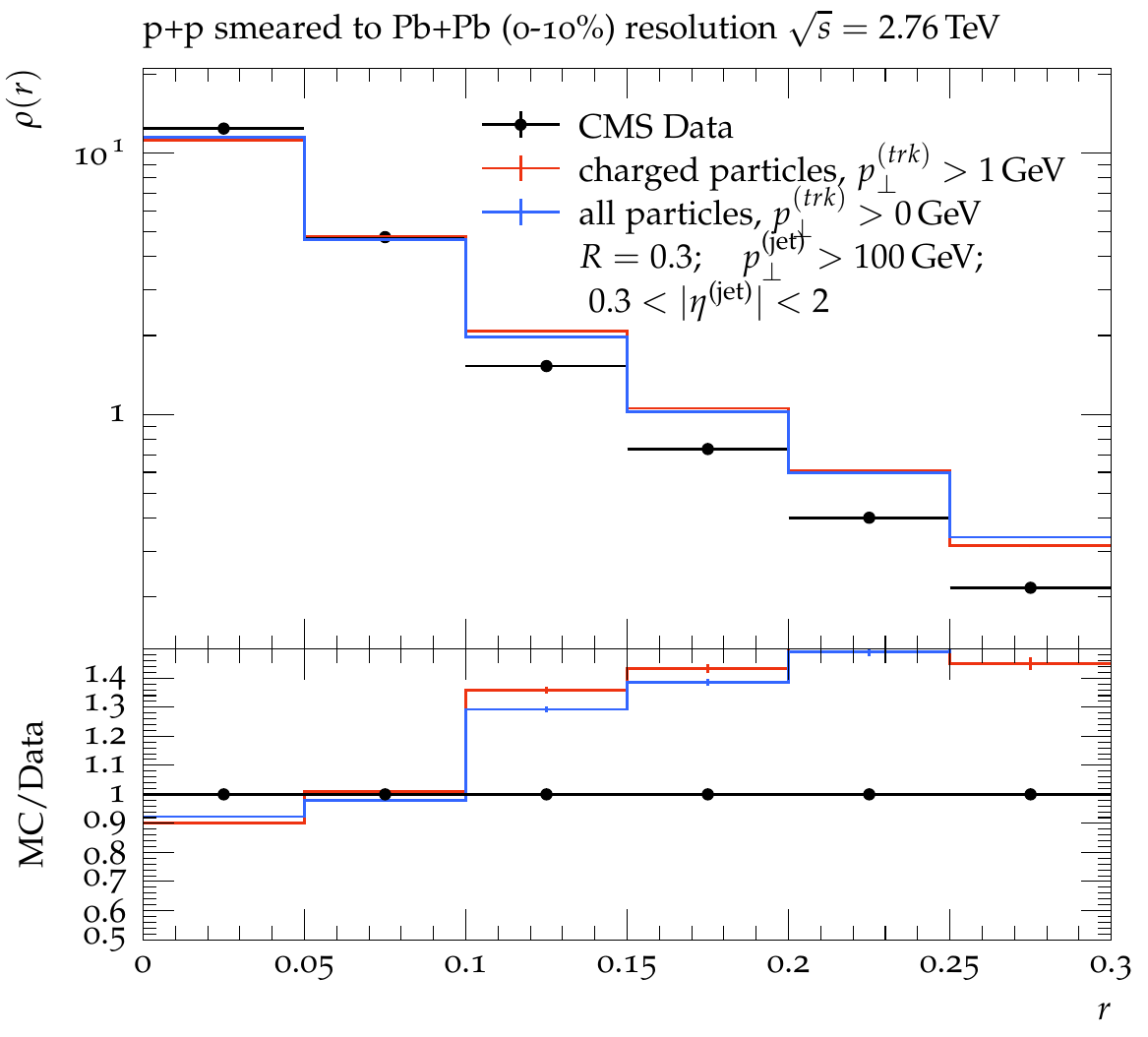}\\
\includegraphics[width=.45\textwidth]{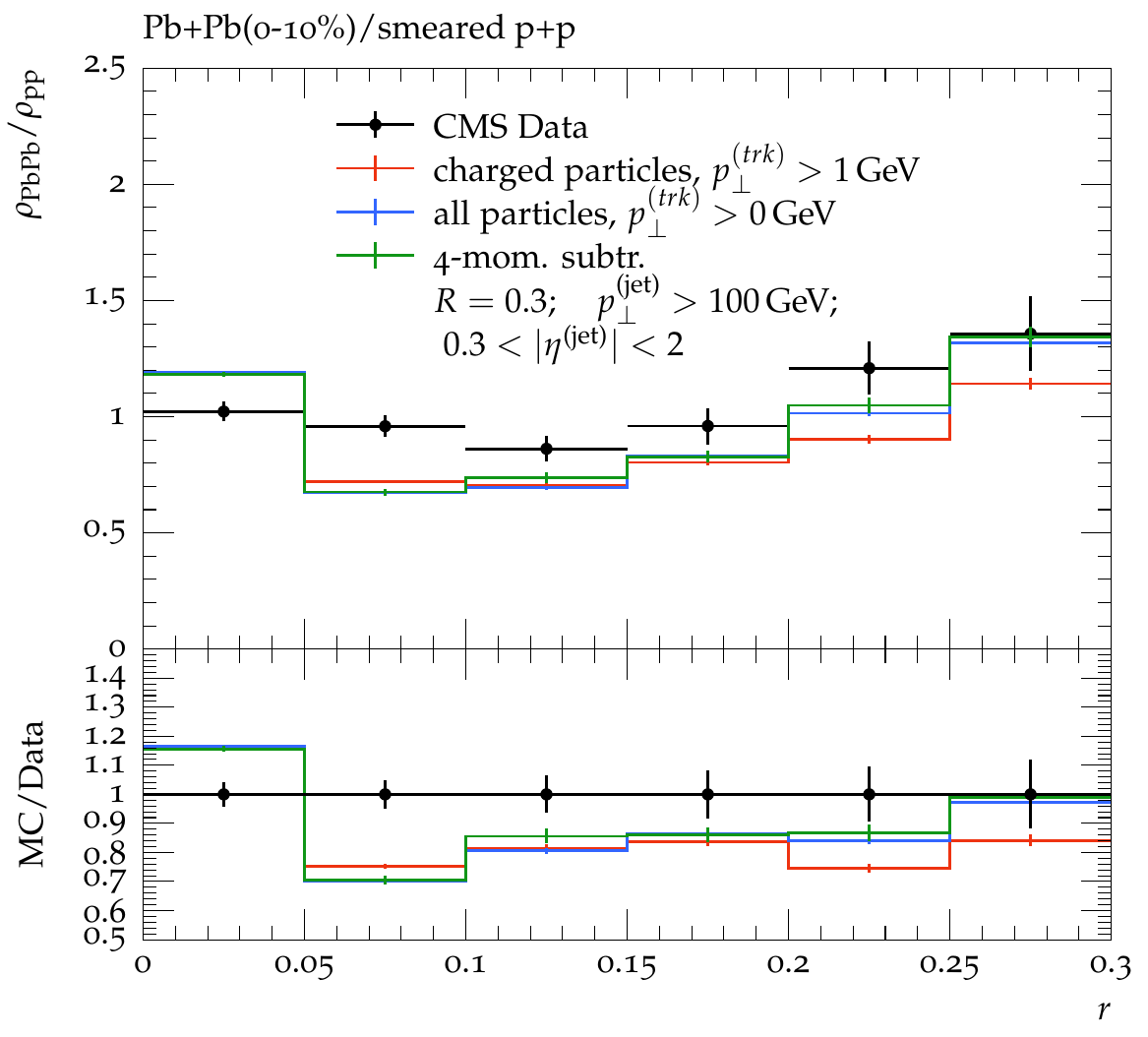}
\caption{Jet - hadron correlation function measured by CMS~\cite{CMS:2013lhm} at $\sqrt{s_\mathrm{NN}} = \unit[2.76]{TeV}$ for $R=0.3$ anti-$\kt$ jets with $\pt^\mathrm{(jet)} > \unit[100]{GeV}$ and $0.3 < |\eta^\mathrm{(jet)}| < 2$. The correlation includes only charged particles with $\pt > \unit[1]{GeV}$. JEWEL+PYTHIA results are shown for four-momentum subtraction and constituent subtraction, the latter including all particles for comparison to four-momentum subtraction and charged particles with $\pt > \unit[1]{GeV}$.}
\label{fig:cmsrho2}
\end{figure}

Figure~\ref{fig:cmsrho2} shows an early measurement of the jet-hadron correlation at $\sqrt{s_\mathrm{NN}} = \unit[2.76]{TeV}$ \cite{CMS:2013lhm}. Here $R=0.3$ jets are reconstructed from both charged and neutral particles with the anti-$\kt$ algorithm and the correlation is taken for charged particles with $\pt > \unit[1]{GeV}$. The measurement is compared to JEWEL+PYTHIA results with constituent subtraction and four-momentum subtraction. Since the data are not unfolded for detector effects, the jet $\pt$ in the MC sample is smeared with the parametrisation from~\cite{CMS:2012ytf}. 
The two subtraction methods agree when all particles are considered in the constituent subtraction case to match what can be done with four-momentum subtraction. When only charged particles above the $\pt$ cut are considered, there are small but significant differences in Pb+Pb while the p+p results remain unchanged. This leads to a small difference in the ratio.

\begin{figure}
\centering
\includegraphics[width=.45\textwidth]{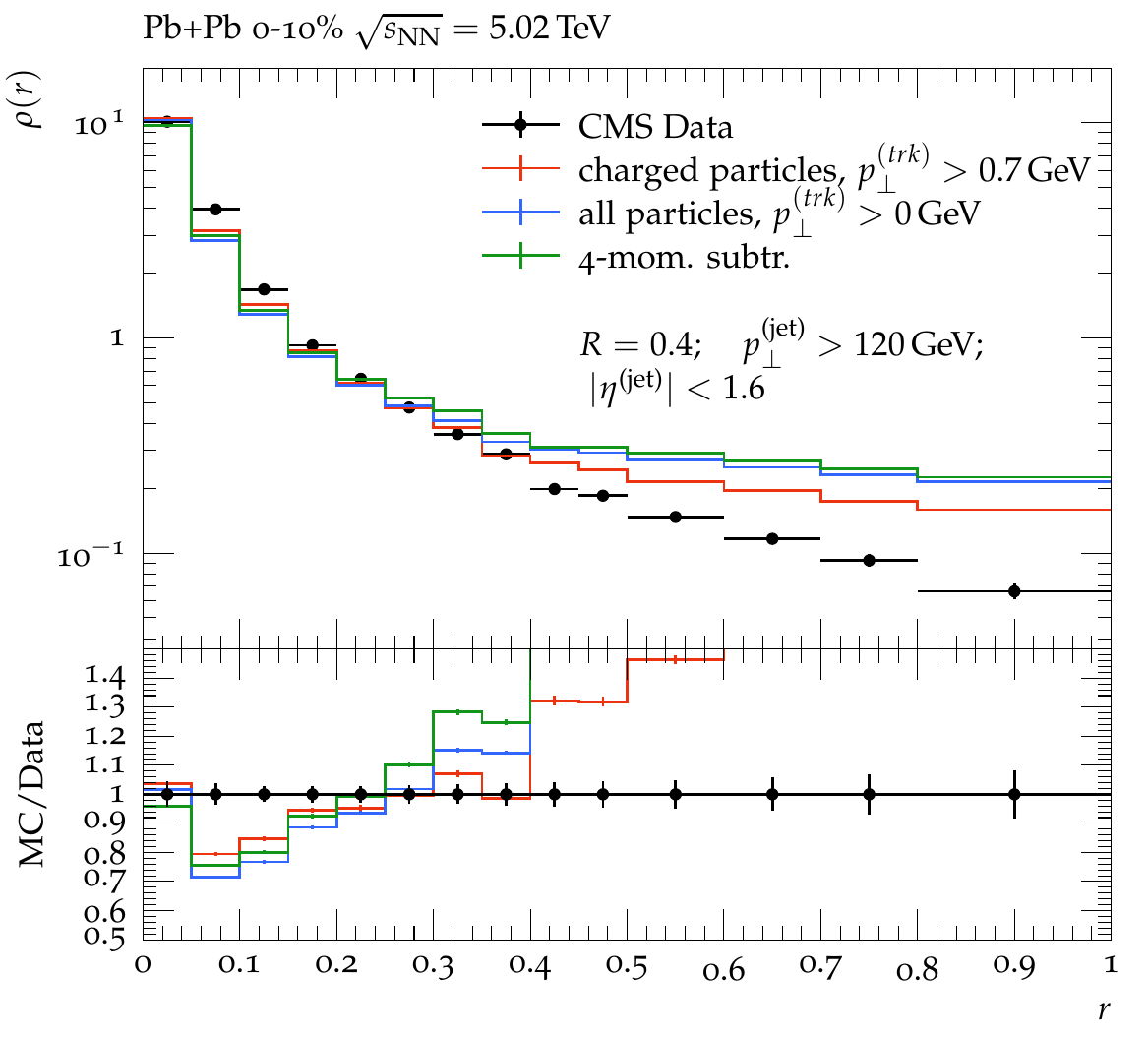}
\includegraphics[width=.45\textwidth]{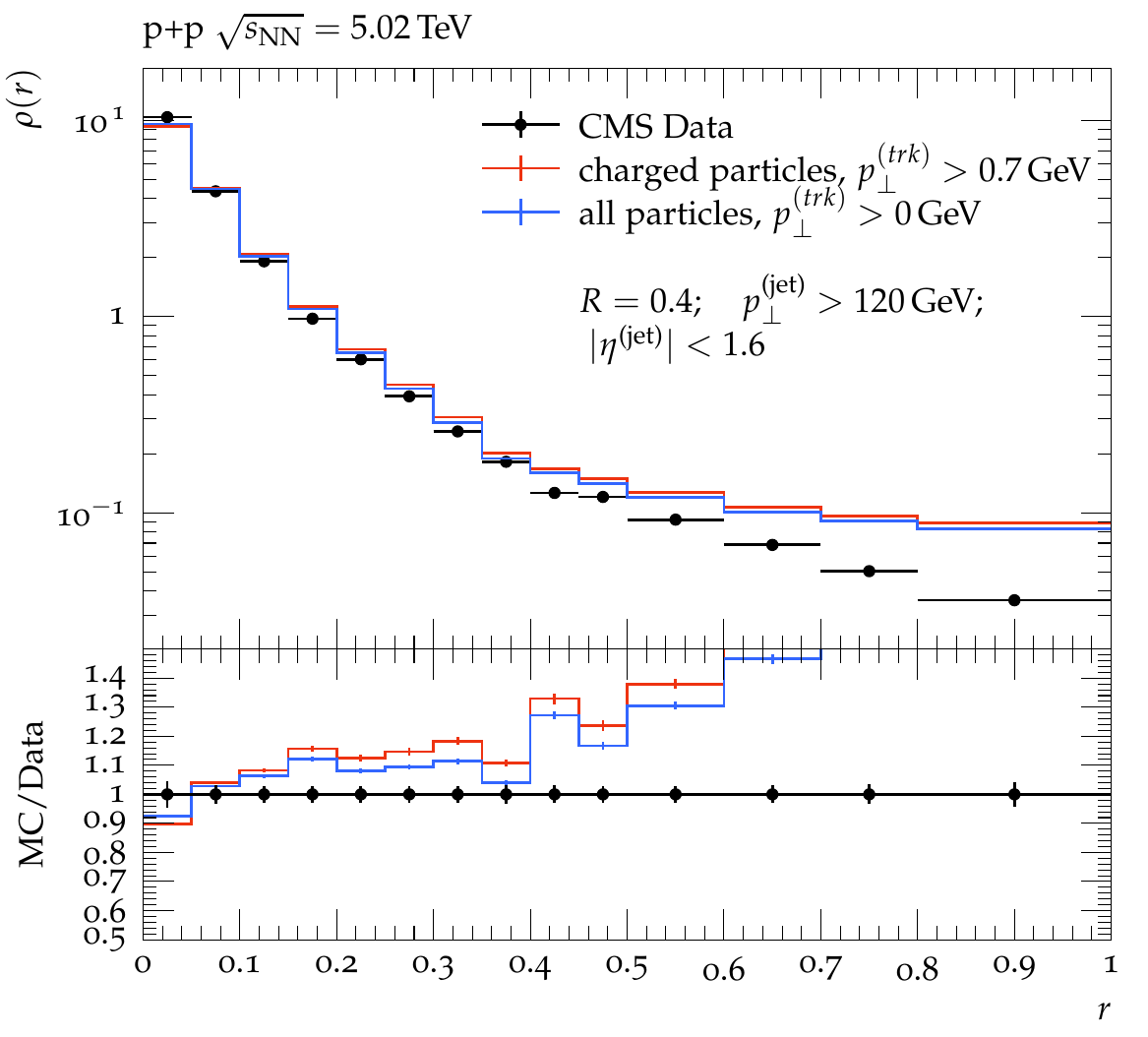}\\
\includegraphics[width=.45\textwidth]{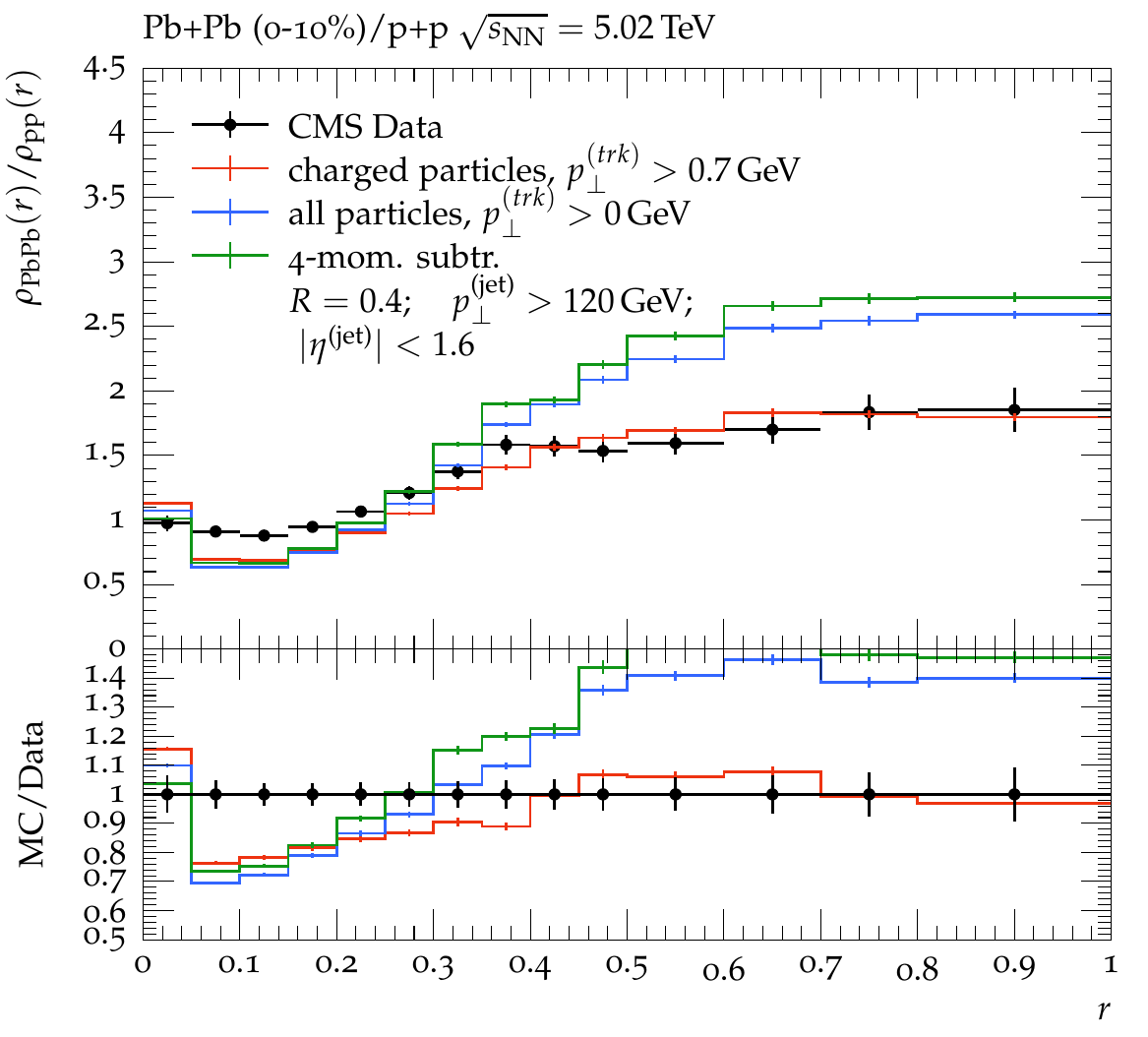}
\caption{Jet - hadron correlation function measured by CMS~\cite{CMS:2018zze} at $\sqrt{s_\mathrm{NN}} = \unit[5.02]{TeV}$ for $R=0.4$ anti-$\kt$ jets with $\pt^\mathrm{(jet)} > \unit[120]{GeV}$ and $|\eta^\mathrm{(jet)}| < 1.6$. The correlation includes only charged particles with $\pt > \unit[0.7]{GeV}$. JEWEL+PYTHIA results are shown for four-momentum subtraction and constituent subtraction, the latter including all particles for comparison to four-momentum subtraction and charged particles with $\pt > \unit[0.7]{GeV}$.
}
\label{fig:cmsrho5}
\end{figure}

In a more recent analysis~\cite{CMS:2018zze} at $\sqrt{s_\mathrm{NN}} = \unit[5.02]{TeV}$ the jet-hadron correlation was extended outside the jet cone (figure~\ref{fig:cmsrho5}). In this measurement $R=0.4$ and charged particles with $\pt > \unit[0.7]{GeV}$ are included in the correlation. The same pattern as in figure~\ref{fig:cmsrho2} is observed: four-momentum and constituent subtraction including all particles yield very similar results, while considering only charged particles above the $\pt$ cut reduces the correlation in Pb+Pb, but not in p+p. The differences are much larger for $r>R$, where the calculation with all particles clearly fails to describe the data while the JEWEL+PYTHIA result for charged particles above the $\pt$ cut nicely reproduces the data for the ratio of the correlations in Pb+Pb and p+p. These findings highlight once more the importance of following the experimental procedures as closely as possible and showcases the greater flexibility of constituent subtraction compared to the old four-momentum subtraction. 

\subsection{Jet fragmentation functions}

\begin{figure}
\centering
\includegraphics[width=.45\textwidth]{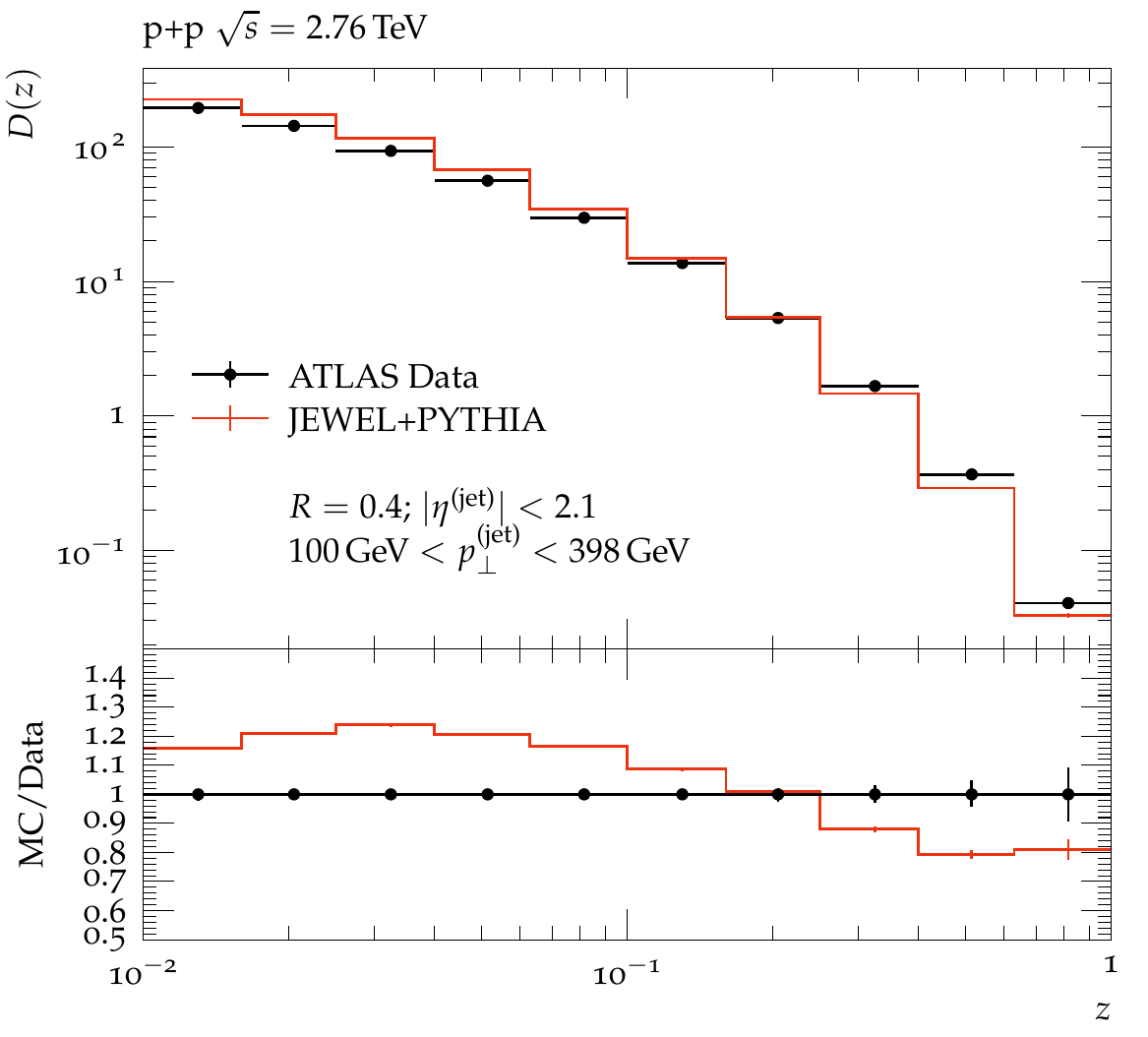}
\includegraphics[width=.45\textwidth]{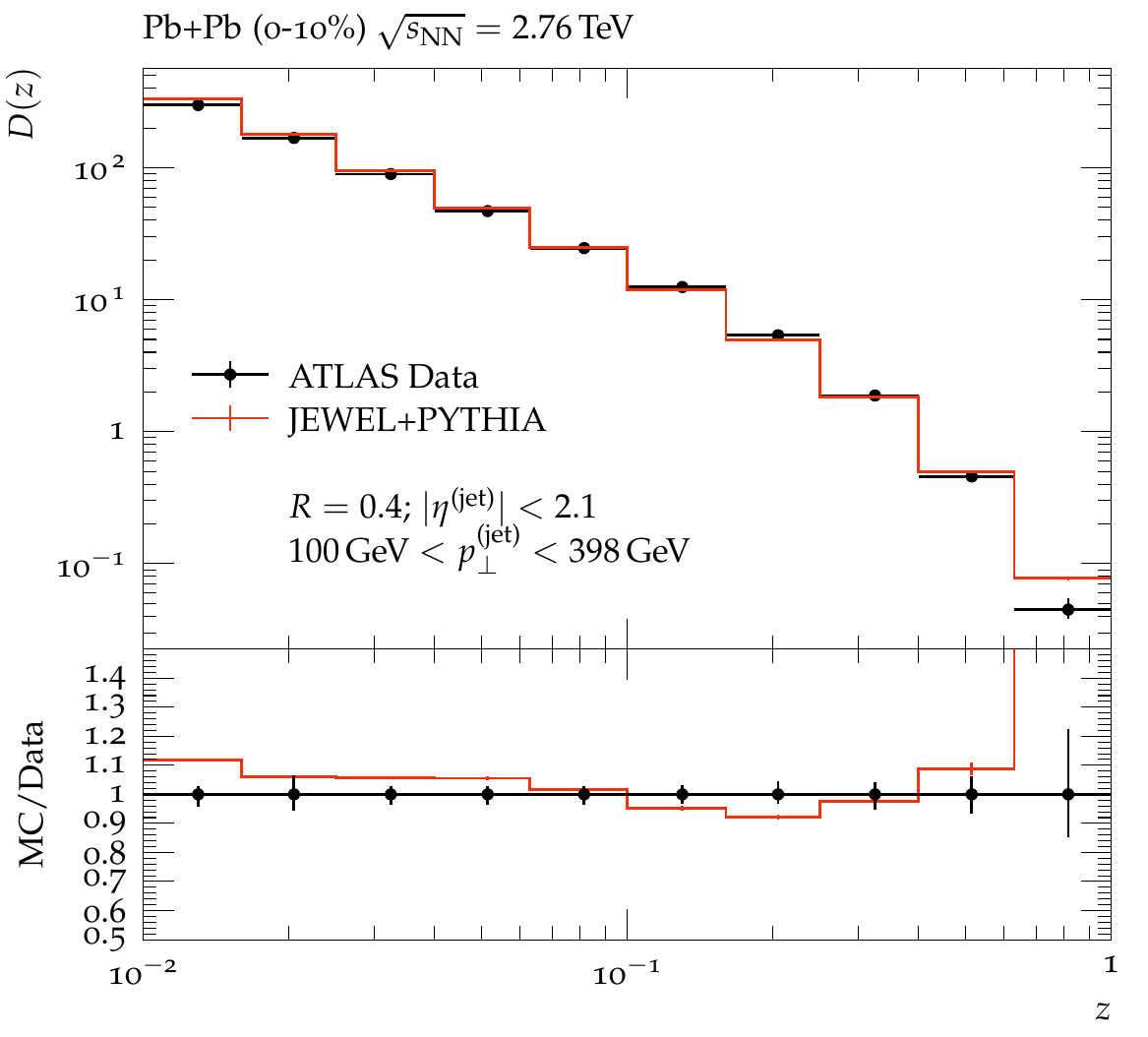}\\
\includegraphics[width=.45\textwidth]{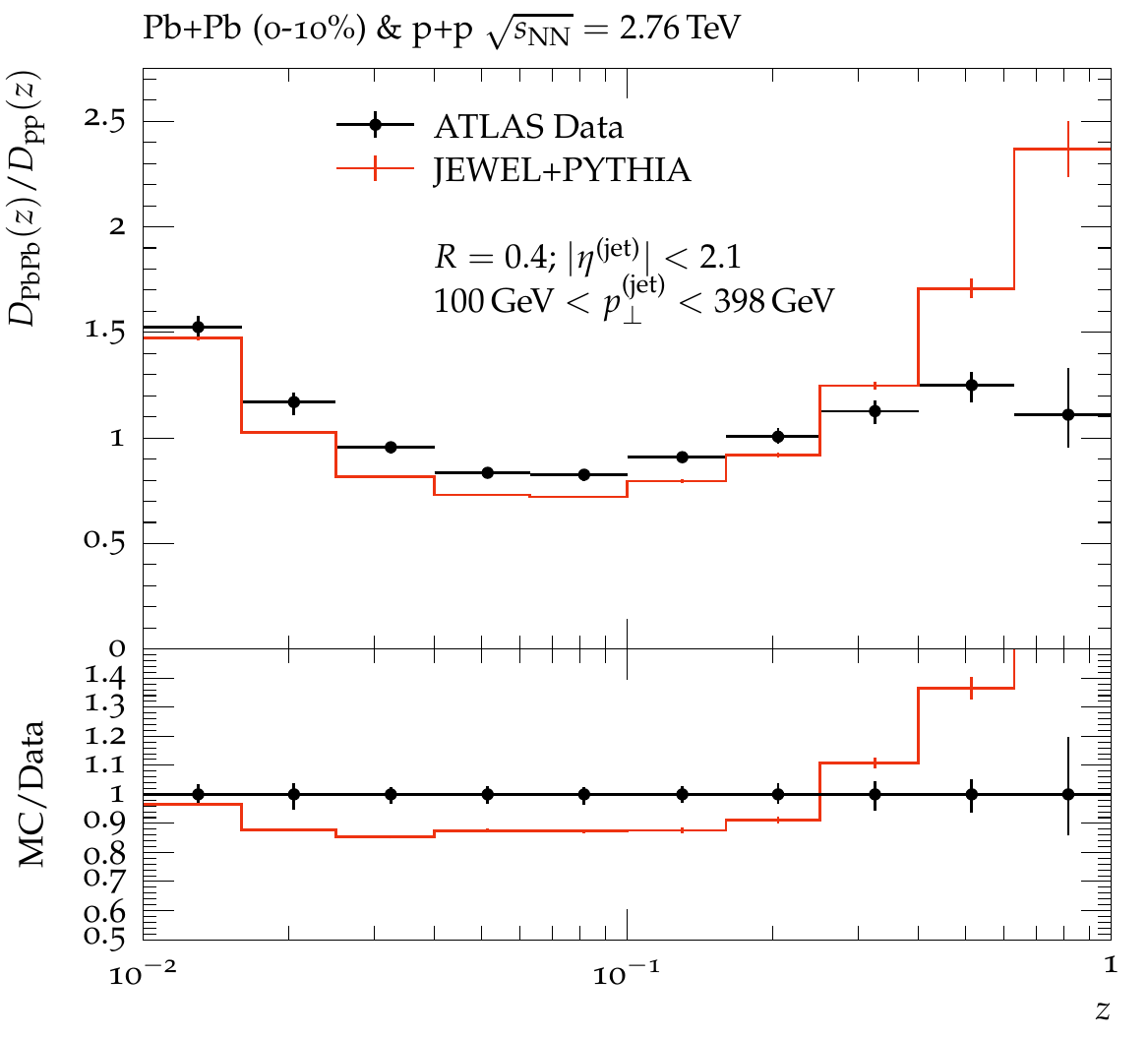}
\caption{Jet fragmentation function for charged particles with $\pt > \unit[1]{GeV}$  as a function of the longitudinal momentum fraction $z$ at $\sqrt{s_\mathrm{NN}} = \unit[2.76]{TeV}$ measured by ATLAS~\cite{ATLAS:2017nre} for $R=0.4$ anti-$\kt$ jets with $\unit[100]{GeV} < \pt^\mathrm{(jet)} < \unit[398]{GeV}$ and $|\eta^\mathrm{(jet)}| < 2.1$ in p+p and Pb+Pb ($\unit[0-10]{\%}$ centrality) collisions. JEWEL+PYTHIA results are shown with constituent subtraction for charged particles with the same $\pt$ cut as in the data.}
\label{fig:atlasFFz}
\end{figure}

\begin{figure}
\centering
\includegraphics[width=.45\textwidth]{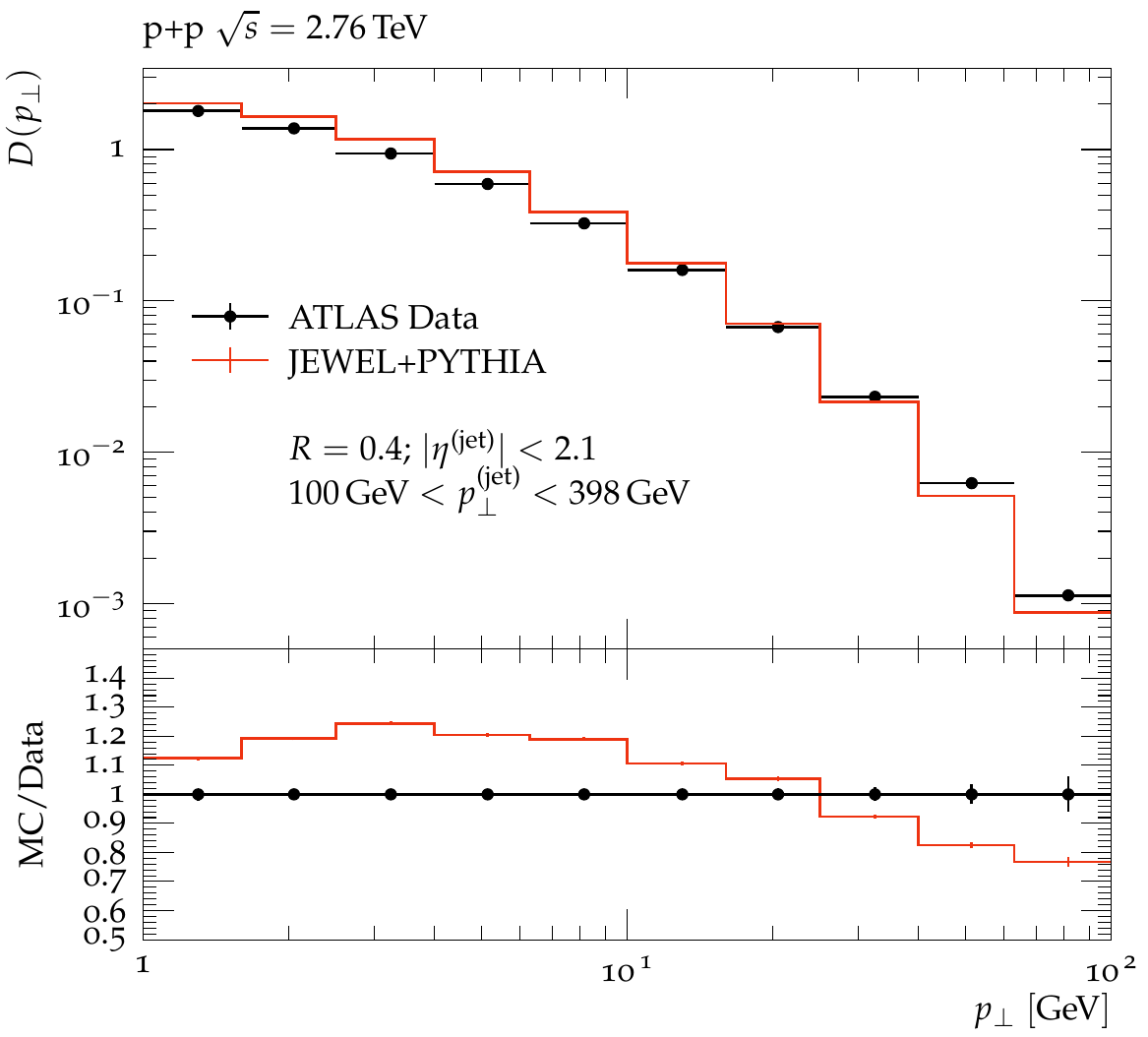}
\includegraphics[width=.45\textwidth]{plots/ATLAS_FF_d01-x01-y01.pdf}\\
\includegraphics[width=.45\textwidth]{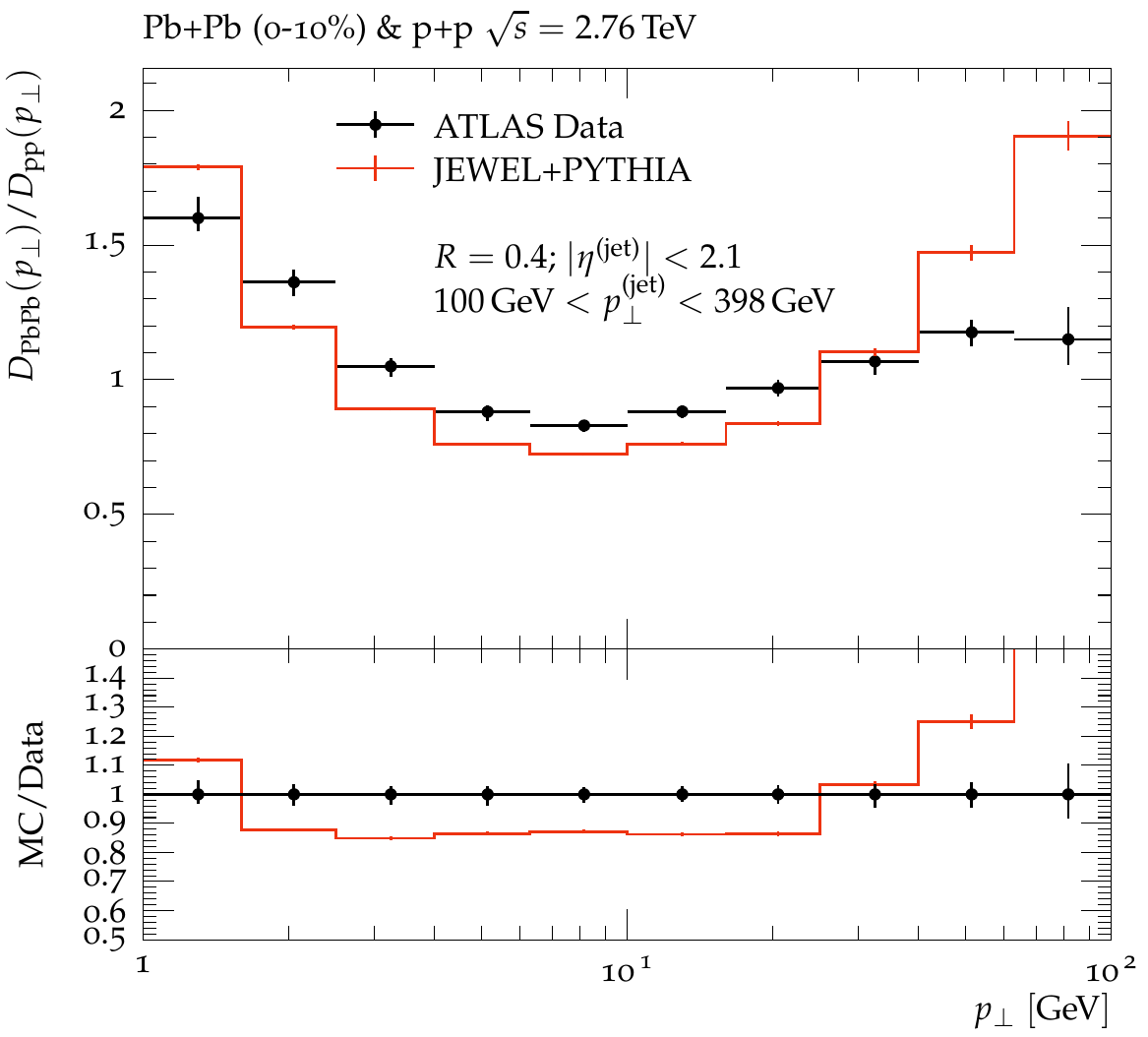}
\caption{Jet fragmentation function for charged particles as a function of particle $\pt$ at $\sqrt{s_\mathrm{NN}} = \unit[2.76]{TeV}$ measured by ATLAS~\cite{ATLAS:2017nre} for $R=0.4$ anti-$\kt$ jets with $\unit[100]{GeV} < \pt^\mathrm{(jet)} < \unit[398]{GeV}$ and $|\eta^\mathrm{(jet)}| < 2.1$ in p+p and Pb+Pb ($\unit[0-10]{\%}$ centrality) collisions. JEWEL+PYTHIA results are shown with constituent subtraction for charged particles.}
\label{fig:atlasFFpt}
\end{figure}

\begin{figure}
\centering
\includegraphics[width=.45\textwidth]{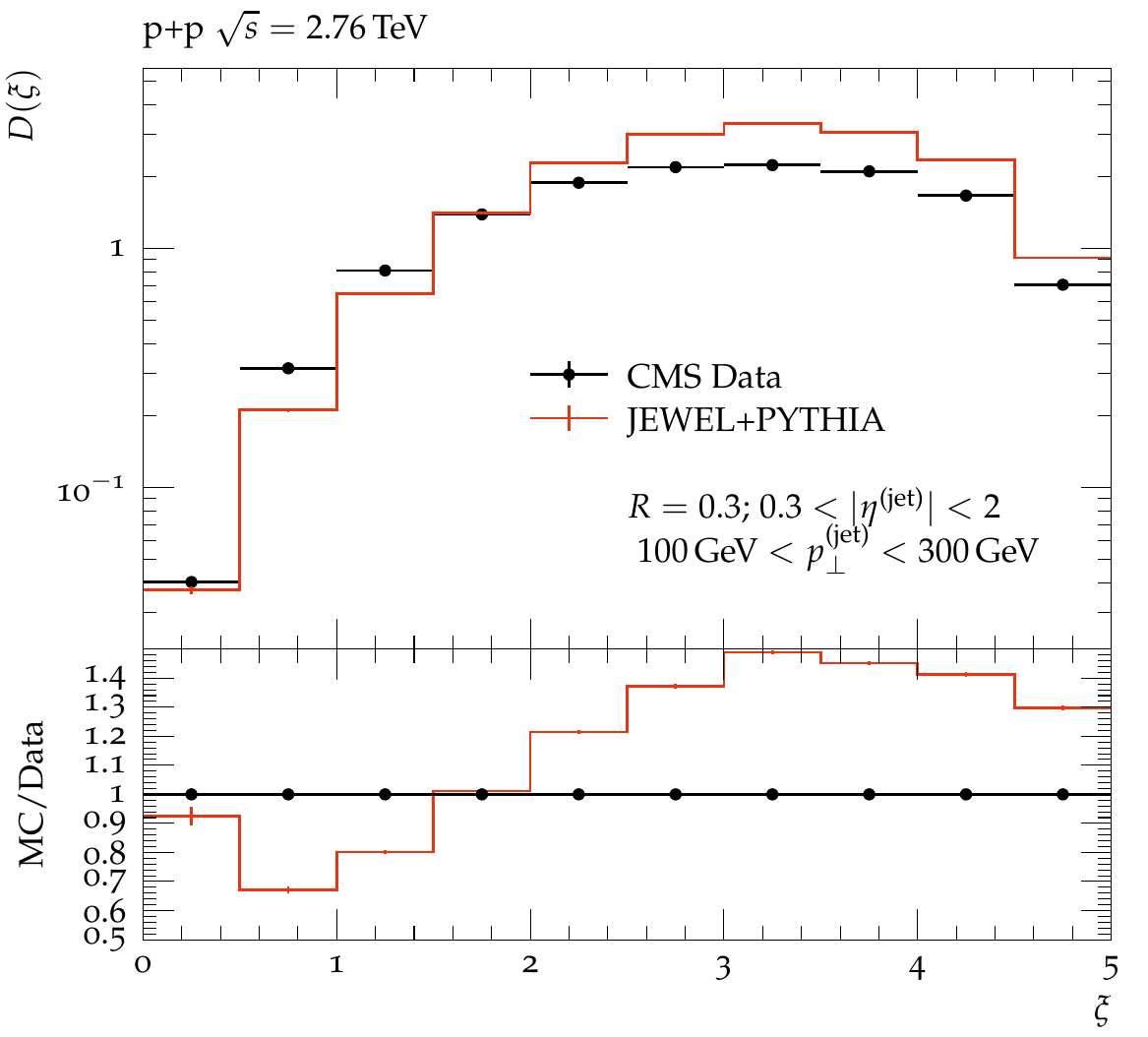}
\includegraphics[width=.45\textwidth]{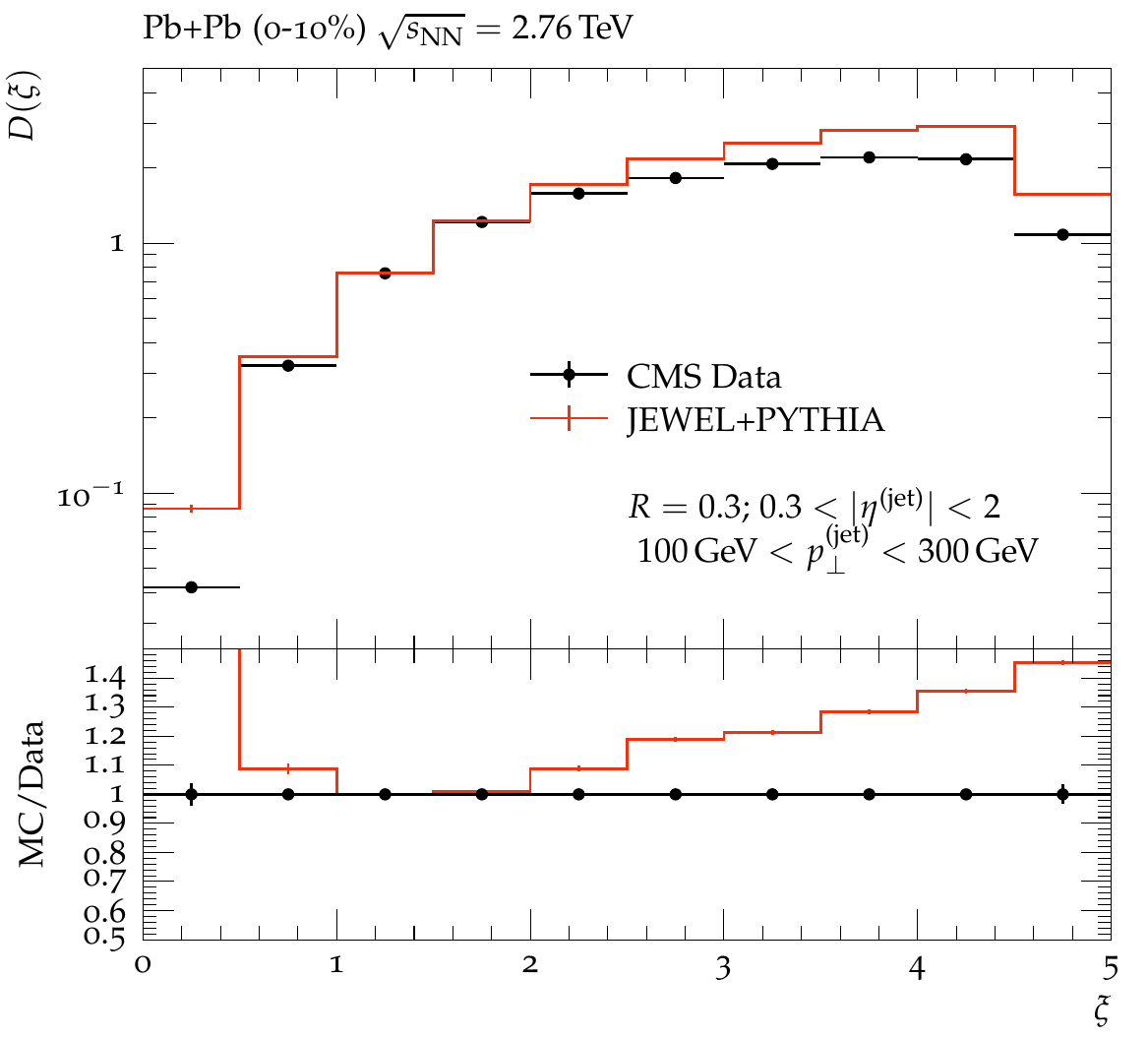}\\
\includegraphics[width=.45\textwidth]{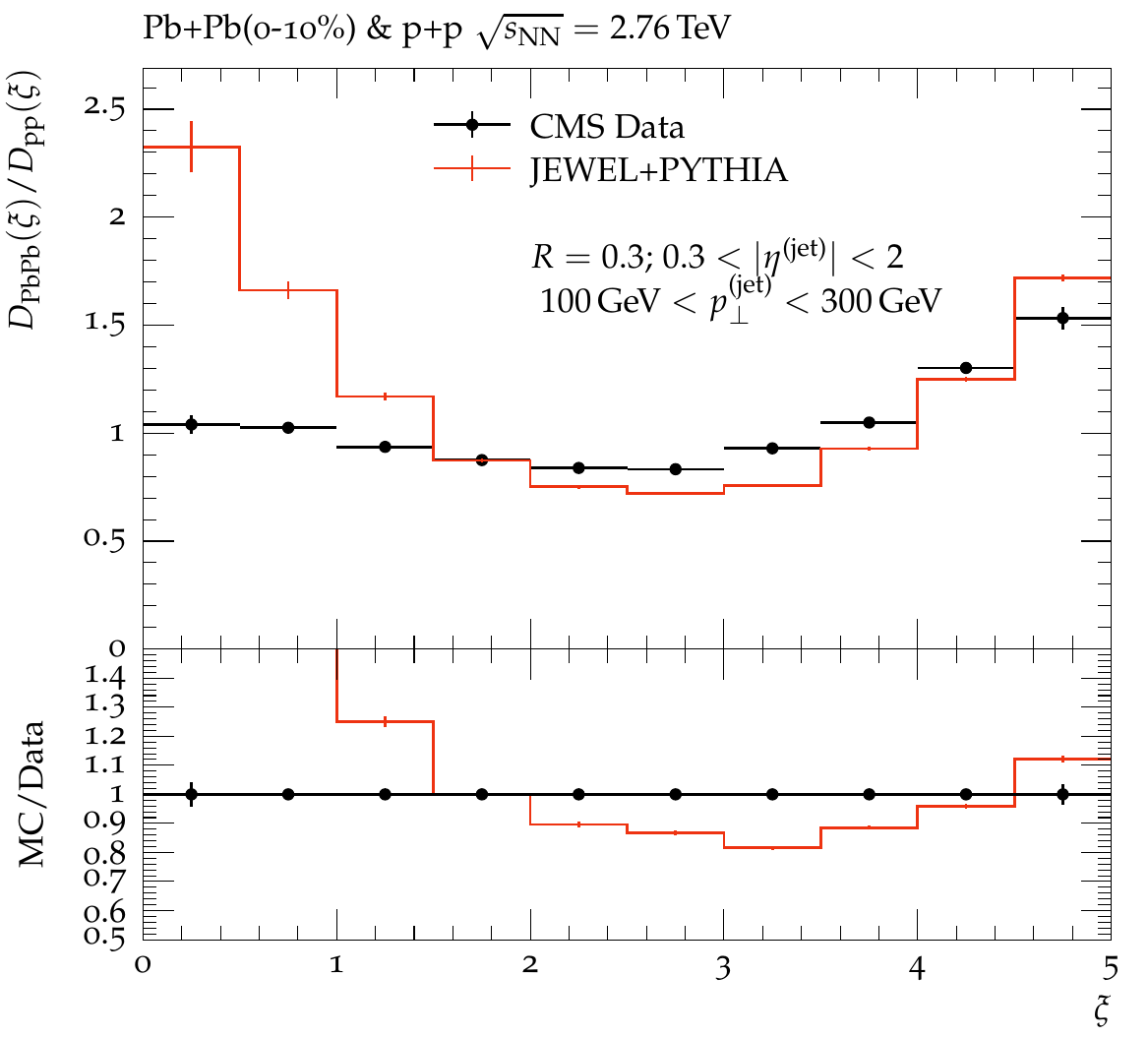}
\caption{Jet fragmentation function for charged particles with $\pt > \unit[1]{GeV}$  as a function of $\xi = \ln(1/z)$ at $\sqrt{s_\mathrm{NN}} = \unit[2.76]{TeV}$ measured by CMS~\cite{CMS:2014jjt} for $R=0.3$ anti-$\kt$ jets with $\unit[100]{GeV} < \pt^\mathrm{(jet)} < \unit[300]{GeV}$ and $0.3 < |\eta^\mathrm{(jet)}| < 2$ in p+p and Pb+Pb ($\unit[0-10]{\%}$ centrality) collisions. JEWEL+PYTHIA results are shown with constituent subtraction for charged particles with the same $\pt$ cut as in the data.}
\label{fig:cmsFFxi}
\end{figure}

\begin{figure}
\centering
\includegraphics[width=.45\textwidth]{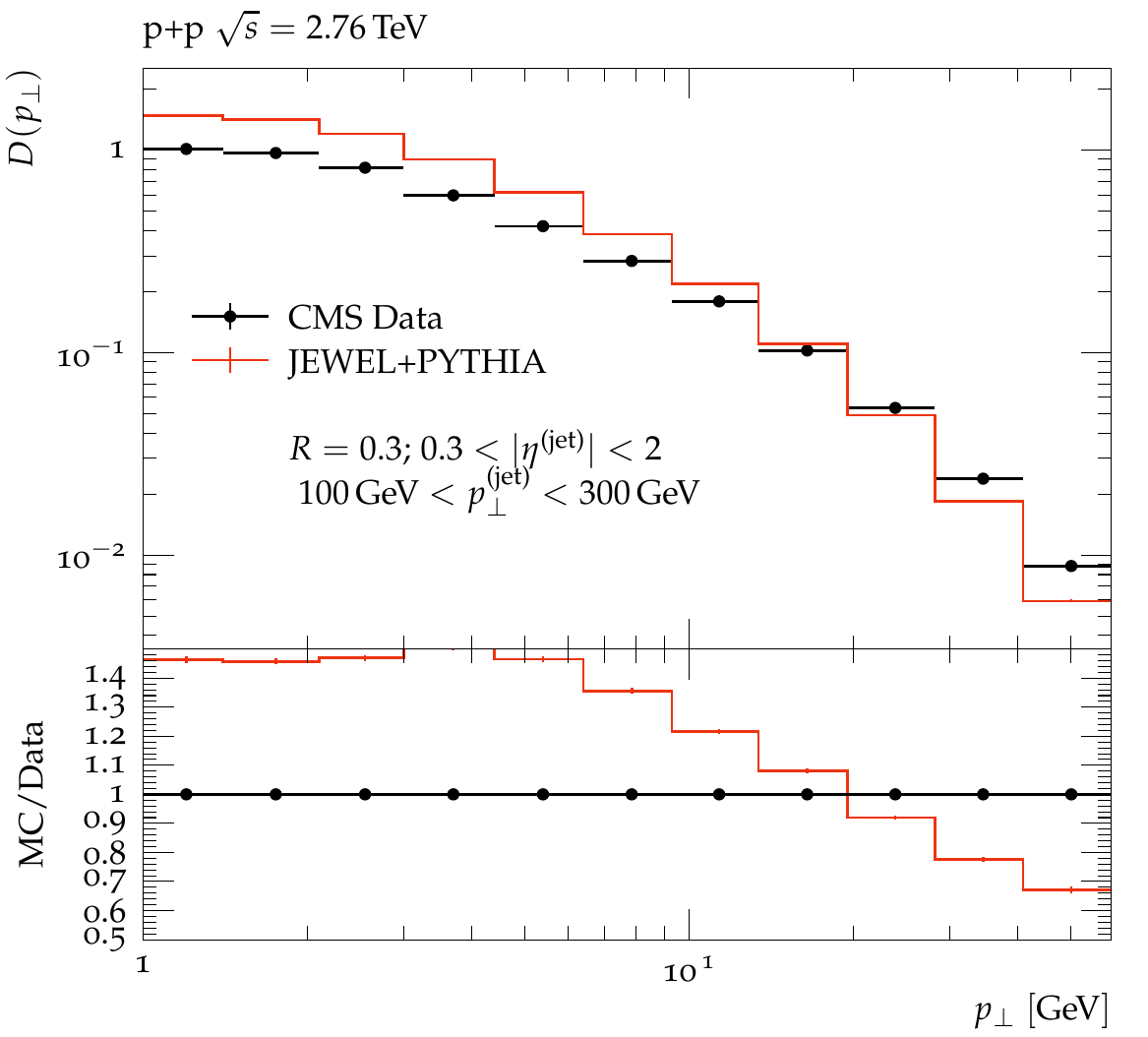}
\includegraphics[width=.45\textwidth]{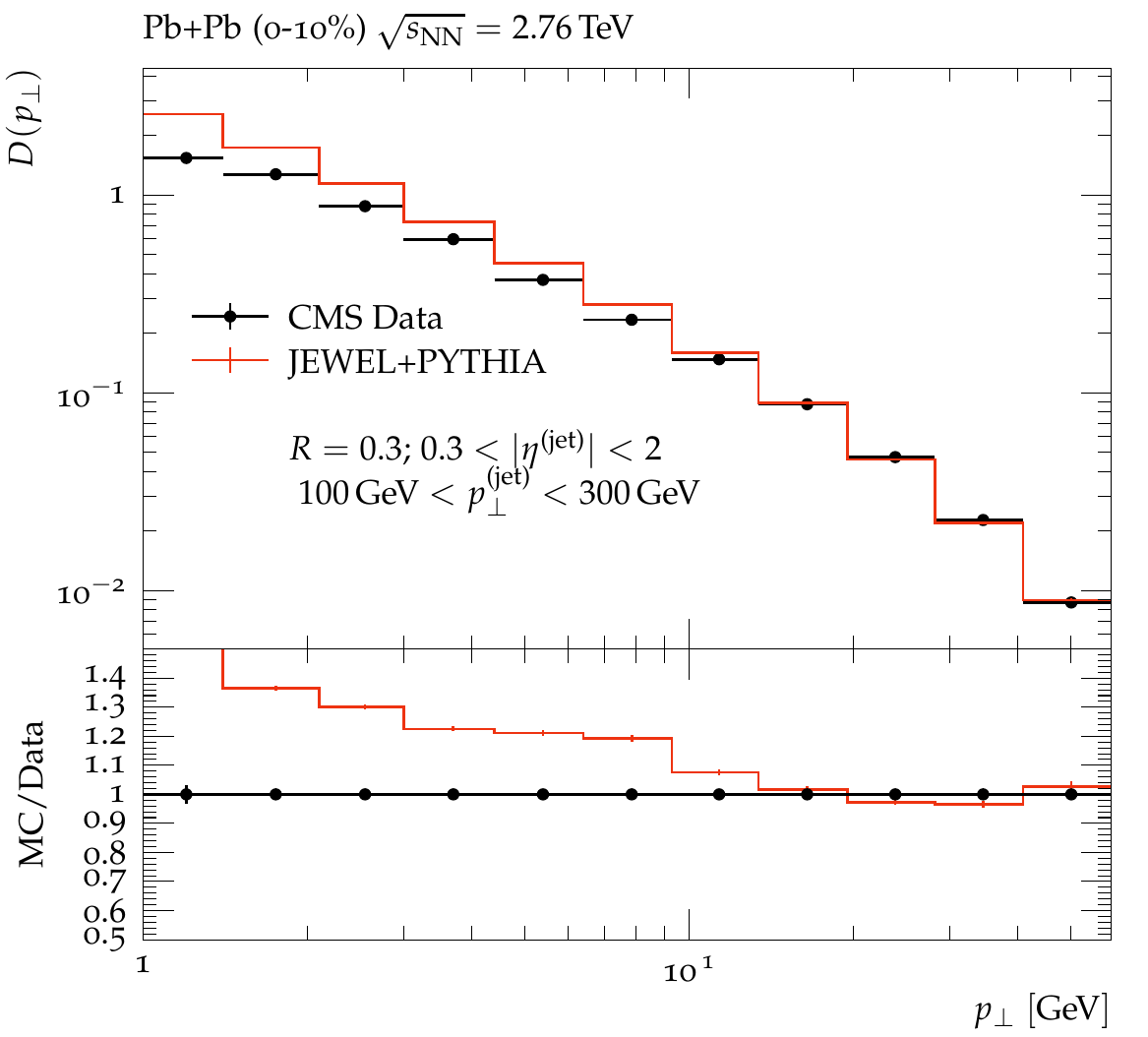}\\
\includegraphics[width=.45\textwidth]{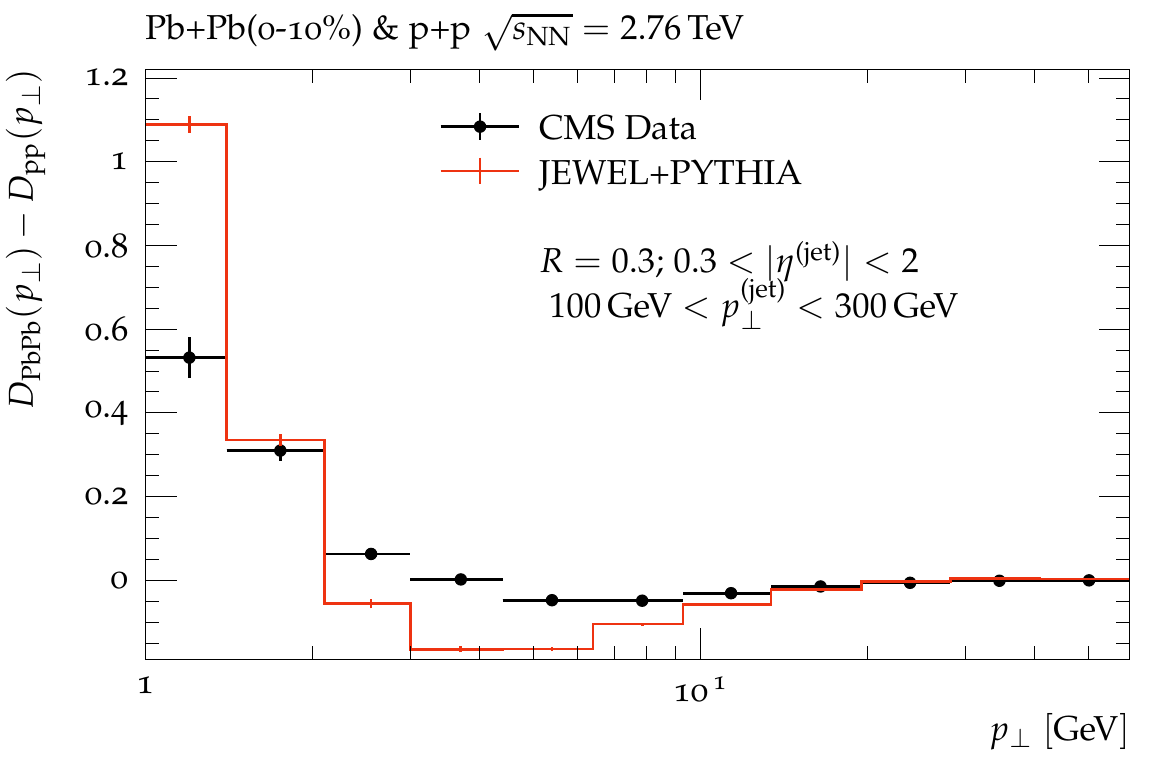}
\caption{Jet fragmentation function for charged particles as a function of particle $\pt$ at $\sqrt{s_\mathrm{NN}} = \unit[2.76]{TeV}$ measured by CMS~\cite{CMS:2014jjt} for $R=0.3$ anti-$\kt$ jets with $\unit[100]{GeV} < \pt^\mathrm{(jet)} < \unit[300]{GeV}$ and $0.3 < |\eta^\mathrm{(jet)}| < 2$ in p+p and Pb+Pb ($\unit[0-10]{\%}$ centrality) collisions. JEWEL+PYTHIA results are shown with constituent subtraction for charged particles.}
\label{fig:cmsFFpt}
\end{figure}

Jet fragmentation functions are an example of an observable that cannot be calculated with four-momentum subtraction in a meaningful way, because it only works at the level of jets and not for hadron distributions\footnote{Some results were shown in~\cite{KunnawalkamElayavalli:2017hxo}, but there only the jet $\pt$ was subtracted while the hadron distribution was not and contained the full recoil contribution. The MC results therefore overshooted the data in the soft part of the fragmentation function.}. It characterises the momentum distribution of charged hadrons found within $\Delta R < R$ of the jet axis in terms of their $\pt$ or the longitudinal momentum fraction
\[ z = \frac{\pt^\mathrm{(track)}}{\pt^\mathrm{(jet)}} \cos \Delta R \quad \text{or} \quad  \xi = \ln(1/z)\,. \]
Figures~\ref{fig:atlasFFz}, \ref{fig:atlasFFpt}, \ref{fig:cmsFFxi} and \ref{fig:cmsFFpt} show a comparison of jet fragmentation function measurements by ATLAS~\cite{ATLAS:2017nre} and CMS~\cite{CMS:2014jjt} at $\sqrt{s_\mathrm{NN}} = \unit[2.76]{TeV}$ to JEWEL+PYTHIA results with constituent subtraction. Thanks to the improved subtraction there is generally a decent agreement between MC and data for small $z$ or $\pt$ (large $\xi$), i.e. in the regime where medium response is important. The hard part of the distribution is generally overestimated by JEWEL+PYTHIA. This is a consequence of a too strong jet collimation effect, that is also seen in other observables (for instance the jet-hadron correlation shown in figures~\ref{fig:cmsrho2} and \ref{fig:cmsrho5}). 
 
\section{Conclusions}
\label{sec:conclusions}
In this paper we have provided a new procedure for background subtraction in events generated with \textsc{Jewel}. 
This procedure, based on constituent subtraction \cite{Berta:2014eza}, 
is more robust that previous procedures applied to \textsc{Jewel} events \cite{KunnawalkamElayavalli:2017hxo}.

Subtracting thermal momenta with constituent subtraction has many advantages over the old four-momentum subtraction method.
This is particularly evident for the jet mass, where results differ substantially for the different methods, with those for constituent subtraction free from artifacts and in clear better agreement with experimental observations.
The main reason for this is that the thermal momenta are subtracted from individual particles instead of (sub)jet four-momenta. This can be done in such a way that the jet mass is much more stable (e.g. the squared mass cannot become negative).

Another, more important, consequence of subtraction from particles is that constituent subtraction can be carried out at event level, before jets are reconstructed. Not only does this lead to less biased jet populations, but it also allows for the introduction of  cuts on the hadron distribution from which observables are calculated. This gives better control over comparisons between theoretical calculations and measurements because the experimental definitions of observables can be followed more closely on the theory side. As seen in section~\ref{sec:results}, this clearly improves the agreement of \textsc{Jewel} calculations with data for the jet mass and jet-hadron correlation at large distances from the jet axis. Other observables, like jet fragmentation functions, which cannot be calculated in a meaningful way using four-momentum subtraction are computed straightforwardly and meaningfully with constituent subtraction.

The results presented here stress the importance of the medium response contributions to the theoretical description of jets in heavy ion collisions with some observables (jet mass, jet profile, fragmentation functions, \ldots) being meaningfully comparable to experimental data only if medium response contributions are included in the theoretical description. \textsc{Jewel} calculations, after background subtraction, should be directly comparable with experimental results where both background subtraction and unfolding for detector effects have been performed.

\acknowledgments

KZ would like to thank Dennis Perepelitsa for an enlightening discussion. 
This work is part of two projects that have received funding from the European Research Council (ERC) under the European Union's Horizon 2020 research and innovation programme: Grant agreement No. 803183, collectiveQCD (KZ) and Grant agreement No. 835105, YoctoLHC (JGM).
JGM further acknowledges financial support from Fundação para a Ciência e a Tecnologia (Portugal) under project CERN/FIS-PAR/0032/2021, he gratefully acknowledges the hospitality of the CERN theory group and thanks Liliana Apolin\'ario for comments on the manuscript.

\bibliographystyle{jhep}
\bibliography{jewelbackground.bib}

\end{document}